\documentclass[11pt,a4paper,british,english,american]{article}
\pdfoutput=1
\usepackage{lmodern}

\usepackage[T1]{fontenc}
\usepackage[latin9]{inputenc}
\usepackage{amsmath}
\usepackage{amssymb}
\usepackage{graphicx}

\makeatletter

\pdfpageheight\paperheight
\pdfpagewidth\paperwidth

\usepackage{jcappub}
\numberwithin{equation}{section}

\pdfoutput=1
\renewcommand\[{\begin{equation}}
\renewcommand\]{\end{equation}}

\@ifundefined{showcaptionsetup}{}{%
 \PassOptionsToPackage{caption=false}{subfig}}
\usepackage{subfig}
\makeatother

\usepackage{babel}
\begin{document}

\title{Unbraiding the Bounce: Superluminality around the Corner }

\subheader{preprint number }

\author[a]{David A. Dobre,}

\author[a]{Andrei V. Frolov,}

\author[a,b]{Jos\'{e} T. G\'{a}lvez Ghersi,}

\author[c]{Sabir Ramazanov,}

\author[c]{and Alexander Vikman}

\affiliation[a]{Department of Physics, Simon Fraser University, }

\affiliation{8888 University Drive, Burnaby, British Columbia V5A 1S6, Canada\\
 }

\affiliation[b]{Perimeter Institute for Theoretical Physics,}

\affiliation{31 Caroline Street North, Waterloo, Ontario, N2L 2Y5, Canada\\
}

\affiliation[c]{CEICO-Central European Institute for Cosmology and Fundamental Physics, }

\affiliation{Institute of Physics of the Czech Academy of Sciences,}

\affiliation{Na Slovance 2, 182 21 Prague 8, Czech Republic\\
}

\emailAdd{ddobre@sfu.ca}

\emailAdd{frolov@sfu.ca}

\emailAdd{joseg@sfu.ca}

\emailAdd{ramazanov@fzu.cz}

\emailAdd{vikman@fzu.cz}

\abstract{We study a particular realization of the cosmological bounce scenario
proposed recently by Ijjas and Steinhardt in \cite{Ijjas:2016tpn}.
First, we find that their bouncing solution starts from a divergent
sound speed and ends with its vanishing. Thus, the solution connects
two strongly coupled configurations. These pathologies are separated
from the bouncing regime by only a few Planck times. We then reveal
the exact structure of the Lagrangian, which reproduces this bouncing
solution. This reconstruction allowed us to consider other cosmological
solutions of the theory and analyze the phase space. In particular,
we find other bouncing solutions and solutions with superluminal sound
speed. These stable superluminal states can be continuously transformed
into the solution constructed by Ijjas and Steinhardt. We discuss
the consequences of this feature for a possible UV-completion. }
\maketitle

\section{Introduction and discussion }

Since 2010 it is well known that minimally-coupled scalar-tensor theories
with \emph{Kinetic Gravity Braiding} \cite{Deffayet:2010qz} can dynamically
violate the Null Energy Condition (NEC)\footnote{For a system with energy-momentum tensor $T^{\mu\nu}$, the NEC holds
provided $T_{\mu\nu}n^{\mu}n^{\nu}\geq0$ for all null / light-like
vectors $n^{\mu}$.} and cross the phantom divide\footnote{This divide is physically impossible to cross \cite{Vikman:2004dc,Hu:2004kh,Caldwell:2005ai,Xia:2007km,Easson:2016klq,deRham:2017aoj}
for k-essence, and theories with minimal coupling to gravity and without
higher derivatives introducing the braiding between the derivatives
of the metric and of the scalar field. } without developing ghost and gradient instabilities\footnote{Despite the positivity of the energy density for linear perturbations,
it is important to note that any theory violating the NEC necessarily
has the energy density unbounded from below \cite{Sawicki:2012pz}. }, see \cite{Deffayet:2010qz,Kobayashi:2010cm,Creminelli:2010ba,Elder:2013gya}.
For a recent review of NEC violation see e.g. \cite{Rubakov:2014jja}.
These \emph{kinetically braided} theories extend the k-essence \cite{ArmendarizPicon:1999rj,ArmendarizPicon:2000ah,ArmendarizPicon:2000dh,Garriga:1999vw}
by a braiding term and the ``decoupling limit of DGP'' \cite{Luty:2003vm,Nicolis:2004qq}
or cubic Galileon \cite{Nicolis:2008in} by higher nonlinear derivative
interactions. The nonlinear interactions break the Galilean symmetry
in the field space even in the Minkowski spacetime, but do not change
the speed of propagation for the gravitational waves. For timelike
field derivatives, \emph{kinetically braided} theories represent imperfect
(super-)fluids with the energy transport along the spatial gradients
of the chemical potential \cite{Pujolas:2011he}. The form of the
braiding term defines the dependence of the transport coefficient
on the chemical potential. In the cubic Galileon the transport coefficient
is given by the square of the chemical potential. In the general \emph{Kinetic
Gravity Braiding,} the transport coefficient can be an arbitrary function
of the chemical potential including the most natural case of a constant. 

After the rediscovery \cite{Deffayet:2011gz,Kobayashi:2011nu} of
the general Horndeski theories \cite{Horndeski:1974} in 2011, these
\emph{kinetically braided} models are often referred to as ``first
two terms of Horndeski theories'' or simply generalized / ``cubic''
Galileons. Since 2011 it is known \cite{Qiu:2011cy,Easson:2011zy}
that these theories can describe spatially-flat Friedmann universes
evolving from contraction to expansion while being manifestly free
of ghost and gradient instabilities around this cosmological bounce.
Hence, these theories can realize a smooth ``healthy'' bounce. The
possibility of the bounce in such systems was briefly mentioned in
\cite{Creminelli:2010ba}. Furthermore, in \cite{Easson:2011zy} it
was demonstrated that one can easily construct spatially-flat ``healthy''
bouncing universes with bouncing solutions of a non-vanishing measure.
This reference established that there is a continuum of such minimally
coupled healthy bouncing theories. The work provided sufficient conditions
in the form of inequalities on the Lagrangian free functions to ensure
a ``healthy'' bounce.\emph{ Kinetically braided} bouncing models
also work in the (unavoidable) presence of normal matter, such as
radiation, etc \cite{Easson:2011zy}. In particular, they allow for
a smooth transition to the radiation dominated époque (``Hot G-Bounce'',
see Fig 1, on page 10 \cite{Easson:2011zy}). In the same work, singularities
in the gravitational metric and in the acoustic metric describing
the cones of propagation for the scalar perturbations, have been discussed.
These singularities can be at the beginning or at the end of the evolution,
or both. Physically, these singularities correspond to (naively infinitely)
strongly coupled configurations, where either the quasi-classical
general relativity (GR) or quasi-classical description of the scalar
field break down. Both theories are not renormalizable, have dynamical
cones of influence / metric, and require a nontrivial ultraviolet
(UV) completion. Under certain assumptions, the presence of these
singularities was later proven for general \emph{kinetically braided}
theories in \cite{Libanov:2016kfc}. This proof was extended to general
Horndeski theories in \cite{Kobayashi:2016xpl} and to theories interacting
with another scalar field in \cite{Kolevatov:2016ppi,Kobayashi:2016xpl}\footnote{It seems that one can have a healthy evolution, without any end or
the beginning singularities, in theories going beyond Horndeski construction
\cite{Creminelli:2016zwa,Cai:2016thi}.}. In this respect, \emph{kinetically braided }theories, as well as
more general Horndeski theories are not that different from regular
GR, where the existence of singularities is a well established fact
\cite{Hawking:1969sw}. The only advantage is that one can relocate
the initial cosmological singularity in the classical dynamical equations
from the expanding to the contracting stage, even in a spatially-flat
Friedmann universe. Hence the big bang could occur not at the beginning
of the cosmological expansion, but at the onset of contraction. This
crucial difference opens up new ways of thinking about initial conditions
in the early universe. This is relevant for the initial conditions
of inflation, which could now be preceded by a contraction, bounce
or even the Minkowski space \cite{Creminelli:2010ba,Creminelli:2012my,Hinterbichler:2012fr,Hinterbichler:2012yn,Pirtskhalava:2014esa,Kobayashi:2015gga,Nishi:2015pta}.
There is plenty of theoretical and philosophical motivation to consider
bounces and NEC violation in the early universe, for recent reviews
see e.g. \cite{Brandenberger:2016vhg,Battefeld:2014uga,Rubakov:2014jja}.
If the NEC can be violated by a physical system one can even consider
such an exotic opportunity as a creation of a universe in a laboratory
\cite{Rubakov:2013kaa}. 

An interesting feature of \emph{kinetically braided} theories is that
the scalar perturbations around generic backgrounds propagate along
an ``acoustic'' cone different from the light cone \cite{Adams:2006sv,Nicolis:2004qq,Deffayet:2010qz,Evslin:2011rj}.
This ``acoustic'' cone can be wider than the light cone or protrude
outside of it just in some directions. In these cases the perturbations
propagate faster than light see e.g. \cite{Adams:2006sv}. Contrary
to k-essence where one can establish subluminality constraints on
the form of the Lagrangian, in \emph{kinetically braided} theories
it seems that there are no such conditions. For a recent discussion
on the dilatationally invariant subclass of these theories see \cite{Kolevatov:2016ppi}.
There are examples \cite{Creminelli:2012my,Easson:2013bda} of such
theories where there is no superluminality for \emph{all} cosmological
configurations, for the proof see \cite{Easson:2013bda}. However,
this only happens in an idealized universe without any external matter.
An unusual property of the\emph{ kinetically braided} theories is
that the value of the sound speed depends not only on the local state
of the field, but also on the energy-momentum tensor of other matter
fields present in the same point of spacetime. Even in the case of
totally subluminal cosmological phase space \cite{Creminelli:2012my,Easson:2013bda},
an introduction of external matter sources instigates the superluminality
at least for some regions of phase space \cite{Easson:2013bda}. 

The superluminality (with hyperbolic equations of motion for perturbations)
per se does not necessarily cause any causal paradoxes, see e.g. \cite{Babichev:2007dw,Geroch:2010da,Bruneton:2006gf,ArmendarizPicon:2005nz,Kang:2007vs,Bruneton:2007si}.
Nevertheless, one can always construct nontrivial non-cosmological
configurations, where closed causal curves (CCC) can be formed at
the level of classical dynamics \cite{Adams:2006sv,Evslin:2011vh,Evslin:2011rj}.
However, similarly to GR where we have \emph{the chronology protection
conjecture} due to Hawking \cite{Hawking:1991nk}, quantum field theory
(QFT) may protect the system from forming CCC in all such theories
with dynamical cones of influence, see e.g. \cite{Babichev:2007dw,Burrage:2011cr,Bruneton:2006gf}. 

On the other hand, there are powerful arguments that effective field
theories (EFT) with at least one configuration permitting superluminality
do not allow for a standard Wilsonian UV-completion in terms of local,
Lorentz-invariant and weakly coupled fields or strings \cite{Adams:2006sv}.
In order to apply these arguments, the superluminal configuration
should belong to the \emph{same} EFT, which UV-completion is investigated.
The latter condition is rather nontrivial, as different semiclassical
states can be isolated by regions with ghosts, regions with strong
coupling, or other features where the EFT breaks down. In these cases
each separated region corresponds to a \emph{different} EFT. The way
out is provided by a recent conjecture concerning a possible Wilsonian
UV-completion in such nonstandard theories. It was conjectured \cite{Dvali:2010jz,Dvali:2010ns}
(see also \cite{Dvali:2016ovn}), that a theory can UV-complete itself
by forming \emph{classicalons} - extended field configurations playing
the role of elementary quantum excitations, hence, the term \emph{classicalization}.
These \emph{classicalons} appear as intermediate long-lived states
and slowly decay into a large number of soft IR elementary excitations.
Later, it was argued that this UV-completion by \emph{classicalization}
takes place, only provided some configurations allow for the superluminal
propagation \cite{Vikman:2012bx,Dvali:2011nj,Dvali:2012zc}. The extended
elementary excitations may induce a non-locality of the UV completion
of these theories \cite{Keltner:2015xda}.\\

Following this discussion, it is expected that general bouncing cosmologies
can only be realized in theories equipped with superluminality around
some configurations. Consequently, these theories cannot be UV-completed
in the standard way. Only classicalization, or maybe some other yet
unknown construction, can UV-complete such bouncing models. \\

In 2016, Ijjas and Steinhardt (IS) proposed in \cite{Ijjas:2016tpn}
an interesting ``inverse'' method. The method allows one to find
particular realizations of the cosmological bounce scenario in a specific
subclass of \emph{kinetically braided} theories. Specifically, they
found a \emph{kinetically braided} model for a given cosmological
evolution $H\left(t\right)$, where $H$ is the Hubble parameter.
\emph{Kinetically braided} theories have two free functions, $K\left(\phi,\partial\phi\right)$
and $G\left(\phi,\partial\phi\right)$. Thus, there is enough freedom
to choose not only $H\left(t\right)$, but also a time-dependence
for one of the two coefficients in the quadratic action for curvature
perturbations. The advantage of this method is that it allows to construct
a theory for a given evolution while keeping a direct control over
perturbations. This procedure enables one to find $\phi\left(t\right)$
and to specify different free functions in the Lagrangian as functions
of time. Hence, this method yields an \emph{implicit} construction
of the Lagrangian of the model realizing the bounce. Using the inverse
method, IS found a particular theory, which accommodates the bounce
free of ghosts and gradient instabilities. For convenience we denote
this realization as \emph{IS-bounce}. The \emph{IS-bounce} was not
only claimed to be free from ghost and gradient instabilities, but
also be exempt from superluminal propagation of perturbations. The
reconstructed solution also included healthy stages before and after
NEC violation. In this way, the system could enter the NEC violating
bouncing stage and leave it without encountering any problem for stability
or UV-completion. These findings are illustrated by explicit numerical
calculations and plots corresponding to two sets of five independent
free parameters. \\

\paragraph{Overview of the paper }

In this paper we analyze the system introduced by IS in \cite{Ijjas:2016tpn}.
First in section (\ref{sec:Model-and-main}) we derive and discuss
main formulas for the dynamics of cosmological solutions and perturbations
in \emph{braided} theories under consideration. In section (\ref{sec:Conformal-Transformations,-Gauge})
we briefly discuss acoustic geometry for cosmological perturbations,
convenient variables and their relations by different gauge and conformal
transformations. Then in section (\ref{sec:Inverse-method}) we uncovered
the \emph{explicit} structure of the Lagrangian by deriving the functions
$k\left(\phi\right)$ and $q\left(\phi\right)$ of the theory, see
equations (\ref{eq:W_IS})-(\ref{eq:q(phi)}). We then used these
results to study the \emph{IS-bounce} for the first set of parameters,
see (\ref{fig:AB}) and (\ref{fig:Speed}). We found that \emph{IS-bounce
}solution starts with a divergent sound speed around $15$  Planck
times, $t_{pl}$, before the NEC violation starts. The sound speed
is still superluminal less than $10\text{ }t_{pl}$ before the onset
of the NEC violation. Moreover, the system enters into the strongly
coupled regime with vanishing sound speed and consequently loses predictive
power in just $15\text{ }t_{pl}$ after the exit from the bouncing
stage with the NEC violation. From the classical perspective the IS
trajectory begins with a singularity of the acoustic metric and ends
in its another singularity. This evolution of the universe is evidently
less appealing, than that during the ``Hot G-Bounce'' scenario mentioned
above. If the system were in a standard weakly coupled vacuum, this
would imply that short-wavelength curvature perturbations evolve from
a state with infinite quantum fluctuations of canonical momentum to
a state with infinite quantum fluctuations of the conjugated canonical
field. Clearly this picture is unphysical, and this implies that the
system is strongly coupled in these singular states. It is a challenge
to modify the theory in such a way that preserves the required evolution,
$H\left(t\right)$, but changes the dynamics on these ultra short
time scales. To make a proper comparison, it is worth noting that
gravity quantum strong coupling scale (and ultimate EFT cutoff) depends
on the number of degrees of freedom $N$ as $M_{pl}/\sqrt{N}$, \cite{Dvali:2007hz},
while already in the Standard Model there are around 100 degrees of
freedom. This implies that the pathologies of the \emph{IS-bounce}
are not really separated in a distinguishable way from the desired
semiclassical evolution.

Further we considered the phase space in this model, see Fig. (\ref{fig:Surface})
and Fig. (\ref{fig:PhaseSpace}). We found other stable bouncing trajectories.
Then we identified stable regions, where superluminality is present.
In parts of these regions the NEC holds, while in others it is violated.
Moreover, there is a region of phase space, where the NEC is broken,
but the sound speed is subluminal. This shows again that the superluminality
is not directly linked to stability of the Phantom stage, c.f. \cite{Dubovsky:2005xd}.
In particular, we found superluminality just around the corner - in
the regions very close to the \emph{IS-bounce. }These regions are
well within the field range corresponding to the NEC violation phase.
Clearly a source or simple interaction can continuously deform these
states into the \emph{IS-bounce }trajectory\emph{. }Hence, these\emph{
}states belong to the same EFT. 

Thus it is impossible to avoid this type of superluminality by modifying
functions $k\left(\phi\right)$ and $q\left(\phi\right)$ in the Lagrangian
outside of the needed field range. In order to attempt escaping superluminality
one has to modify either the desired evolution $H\left(t\right)$
or the structure of the theory or both. 

Other interesting findings include the following. The \emph{IS-bounce}
is a separatrix, see Fig. (\ref{fig:PhaseSpace}). This solution goes
through the singularity of the equation of motion, similarly to the
singular trajectories found in \cite{Vikman:2004dc}. For the second
choice of the parameters used in \cite{Ijjas:2016tpn} to obtain their
Fig. 3, we could not reproduce the IS claims. It seems that below
Fig. 3 from \cite{Ijjas:2016tpn} there is a typo somewhere either
in the set of parameters or in the form of the functions. \\

To conclude, we think it is interesting to understand the consequences
of the possible bounces in the early universe. Though, so far, this
nonstandard option for the early universe seems to be inseparable
from superluminality and a nonstandard UV-completion with \emph{classicalization}
as the only current candidate for the latter. 

\section{Model and main equations\label{sec:Model-and-main} }

The \emph{IS-bounce} uses a class of Kinetic Gravity Braiding theories
with explicitly strongly \emph{broken} shift-symmetry $\phi\rightarrow\phi+c$
\begin{equation}
S=\frac{1}{2}\int d^{4}x\,\sqrt{-g}\left(k\left(\phi\right)\left(\partial\phi\right)^{2}+\frac{1}{2}q\left(\phi\right)\left(\partial\phi\right)^{4}+\left(\partial\phi\right)^{2}\Box\phi\right)\,,\label{eq:IS_Lagrangian}
\end{equation}
where 
\begin{equation}
\left(\partial\phi\right)^{2}\equiv g^{\mu\nu}\partial_{\mu}\phi\partial_{\nu}\phi\equiv2X\,,\qquad\Box\phi\equiv g^{\mu\nu}\nabla_{\mu}\nabla_{\nu}\phi\,,\label{eq:X}
\end{equation}
where $\nabla_{\mu}$ is the usual Levi-Civita connection \foreignlanguage{english}{}\footnote{Further we use: the standard notation $\sqrt{-g}\equiv\sqrt{-\text{det}g_{\mu\nu}}$
where $g_{\mu\nu}$ is the metric, the signature convention $\left(+,-,-,-\right)$
(contrary to \cite{Ijjas:2016tpn}), and the units $c=\hbar=1$, $M_{\text{Pl}}=\left(8\pi G_{\text{N}}\right)^{-1/2}=1$. }. The scalar field is supposed to be minimally coupled to gravity.
Hence, the theory is defined by two free functions $k\left(\phi\right)$
and $q\left(\phi\right)$. In notation of \cite{Deffayet:2010qz}
where generic theories of the type 
\begin{equation}
S=\int d^{4}x\,\sqrt{-g}\left[K\left(X,\phi\right)+G\left(X,\phi\right)\Box\phi\right]\,,\label{eq:general_action}
\end{equation}
were introduced, we have\footnote{At the beginning the authors of \cite{Ijjas:2016tpn} also used $G\left(X,\phi\right)=b\left(\phi\right)X$,
however this additional free function $b\left(\phi\right)$ can be
eliminated by the simple field-redefinition: $d\bar{\phi}=b^{-1/3}\left(\phi\right)d\phi$. } 
\begin{equation}
K\left(X,\phi\right)=k\left(\phi\right)X+q\left(\phi\right)X^{2}\,,\qquad G\left(X,\phi\right)=X\,.\label{eq:KGB_identifiaction}
\end{equation}
This identification allows us to directly use all necessary formulas
derived in \cite{Deffayet:2010qz} (see also \cite{Kobayashi:2010cm})
for \emph{arbitrary} $K\left(X,\phi\right)$ and $G\left(X,\phi\right)$
for the background dynamics and perturbations in the spatially-flat
Friedmann universe
\[
ds^{2}=dt^{2}-a^{2}\left(t\right)d\mathbf{x}^{2}\text{ .}
\]
 Below, instead of rederiving formulas for the particular case (\ref{eq:KGB_identifiaction}),
as it was done by IS, we use general results from \cite{Deffayet:2010qz}.
In particular, the pressure is 
\begin{equation}
\mathcal{P}\left(\phi,X,\ddot{\phi}\right)=K-2XG_{,\phi}-2XG_{,X}\ddot{\phi}=kX+qX^{2}-2X\ddot{\phi}\text{ ,}\label{eq:pressure}
\end{equation}
the charge density with respect to shifts $\phi\rightarrow\phi+c$
\begin{equation}
J=\dot{\phi}\left(K_{,X}-2G_{,\phi}+3\dot{\phi}HG_{,X}\right)=\dot{\phi}\left(k+2qX+3\dot{\phi}H\right)\text{ ,}\label{eq:charge_density}
\end{equation}
where $H=\dot{a}/a$ is the Hubble parameter. The variation of the
action (\ref{eq:general_action}) with respect to the field $\phi$
gives an equation of motion, which in terms of the charge density
takes the following elegant form 
\begin{equation}
\dot{J}+3HJ=\mathcal{P}_{,\phi}\text{ .}\label{eq:EoM_J}
\end{equation}
The \emph{kinetic braiding} with gravity reveals itself in the presence
of the $\dot{H}$ in this equation. 

The general expression for the energy density 
\begin{equation}
\varepsilon\left(\phi,\dot{\phi},H\right)=\dot{\phi}J-\mathcal{P}+2XG_{,X}\ddot{\phi}=2X\left(K_{,X}-G_{,\phi}+3\dot{\phi}HG_{,X}\right)-K\text{ ,}\label{eq:energy_desnity}
\end{equation}
reduces for the choice (\ref{eq:KGB_identifiaction}) to
\begin{equation}
\varepsilon=kX+3qX^{2}+6\dot{\phi}HX\text{ .}\label{eq:energy_density_IS}
\end{equation}
Further the first Friedmann equation reads 
\begin{equation}
3H^{2}=\varepsilon=kX+3qX^{2}+6\dot{\phi}HX\text{ ,}\label{eq:First_Friedmann}
\end{equation}
while for the second equation we have 
\begin{equation}
\dot{H}=-\frac{1}{2}\left(\varepsilon+\mathcal{P}\right)=XG_{,X}\ddot{\phi}-\frac{1}{2}\dot{\phi}J\text{ .}\label{eq:Second_Friedmann}
\end{equation}
It is important that the energy density and the first Friedmann equation
contain a term linear in $H$. Therefore for \emph{kinetically braided}
systems the branches resulting from the first Friedmann equation do
not correspond to expansion and contraction of the universe. Indeed,
solving the quadratic equation we obtain 
\begin{equation}
H_{\pm}=XG_{,X}\dot{\phi}\pm\sqrt{\left(XG_{,X}\dot{\phi}\right)^{2}+\frac{1}{3}\left(2X\left(K_{,X}-G_{,\phi}\right)-K\right)}\text{ .}\label{eq:Friedmann_with_branches}
\end{equation}
This relation implies that not all configurations $\left(\phi,\dot{\phi}\right)$
with positive energy density are allowed, but only those satisfying
an additional condition 
\begin{equation}
6X\left(XG_{,X}\right)^{2}+2X\left(K_{,X}-G_{,\phi}\right)-K\geq0\text{ .}\label{eq:condition}
\end{equation}
Finally it is convenient to write an equation of motion for the scalar
field (\ref{eq:EoM_J}) where $\dot{H}$ is expressed through (\ref{eq:Second_Friedmann})
\begin{equation}
D\ddot{\phi}+3J\left(H-\dot{\phi}XG_{,X}\right)+\varepsilon_{,\phi}=0\text{ ,}\label{eq:EoM_phi}
\end{equation}
where general expression 
\begin{equation}
D=K_{,X}+2XK_{,XX}-2G_{,\phi}-2XG_{,X\phi}+6\dot{\phi}H\left(G_{,X}+XG_{,XX}\right)+6X^{2}G_{,X}^{2}\,,\label{eq:D_for_ghosts}
\end{equation}
reduces for the choice (\ref{eq:KGB_identifiaction}) to 
\begin{equation}
D=k+6Xq+6\dot{\phi}H+6X^{2}\,.\label{eq:D_IS-1}
\end{equation}

Now we are prepared to write the formulas for the perturbations. We
use the ``unitary'' gauge, and embed the spacelike hypersurface
$\delta\phi=0$ into the perturbed Friedmann universe using the ADM
decomposition 
\begin{equation}
ds^{2}=N^{2}dt^{2}-a^{2}e^{2\mathcal{R}}\delta_{ik}\left(N^{i}dt+dx^{i}\right)\left(N^{k}dt+dx^{k}\right)\text{ ,}\label{eq:perturbed_line_R}
\end{equation}
which is useful to compare with the most general line element with
scalar perturbations written in conformal time $\tau$ 
\begin{equation}
ds^{2}=a^{2}\left[\left(1+2\varphi\right)d\tau^{2}+2B_{,i}dx^{i}d\tau-\left[\left(1-2\psi\right)\delta_{ik}-2E_{,ik}\right]dx^{i}dx^{k}\right]\text{ .}\label{eq:standard_perturbations}
\end{equation}
As it was pointed out in \cite{Kobayashi:2010cm}, the variable $\mathcal{R}$
is not a \emph{comoving} curvature perturbation, because there is
an energy flow $T_{i}^{t}=\dot{\phi}^{3}G_{X}\partial_{i}\delta N$
in the gauge (\ref{eq:perturbed_line_R}). Expressing the lapse fluctuation
$\delta N$ through the longitudinal part of the momentum constraint
(see (A.5) on page 36 \cite{Deffayet:2010qz}) 
\begin{equation}
\left(H-\dot{\phi}XG_{,X}\right)\delta N=\dot{\mathcal{R}}\text{ ,}\label{eq:constraint_momentum}
\end{equation}
one obtains the quadratic action for curvature perturbations $\mathcal{R}$,
\footnote{Note that \cite{Ijjas:2016tpn} does not use the canonical normalization
with $1/2$ in front of the action. Hence our coefficients $A\left(t\right)$
and $B\left(t\right)$ are twice larger then those in \cite{Ijjas:2016tpn}. } (see (A.8) on page 36 \cite{Deffayet:2010qz}) 
\begin{equation}
S_{c}=\frac{1}{2}\int dt\,d^{3}x\,a^{3}\left(A\left(t\right)\dot{\mathcal{R}}^{2}-\frac{B\left(t\right)}{a^{2}}\left(\partial_{i}\mathcal{R}\right)^{2}\right)\,.\label{eq:quadratic_action_scalars}
\end{equation}
The formula for the normalization of the kinetic term is given by
(A.9) page 36, \cite{Deffayet:2010qz}
\begin{equation}
A=\frac{2XD}{\left(H-\dot{\phi}XG_{,X}\right)^{2}}\,.\label{eq:Our_Normalization}
\end{equation}
Hence, it is the coefficient $D$, given by (\ref{eq:D_for_ghosts}),
in front of the second derivative in the reduced equation of motion
(\ref{eq:EoM_phi}) which determines ($D>0$) whether the perturbations
are free of ghosts. It is interesting to note that curves on phase
space with $D=0$ correspond to an infinitely-strong coupling of perturbations
and to pressure-like curvature singularity \cite{Deffayet:2010qz},
where GR breaks down, see (\ref{eq:pressure}) and (\ref{eq:EoM_phi}).
Clearly the reduced equation of motion is singular on these curves.
The quantity $D$ corresponds to the determinant of the matrix in
front of the second derivatives $\left(\ddot{\phi},\ddot{a}\right)$
in the equations (\ref{eq:Second_Friedmann}) and (\ref{eq:EoM_J}). 

The sound speed is given by the formula (A.11) page 36, \cite{Deffayet:2010qz}
\begin{equation}
c_{s}^{2}=\frac{B\left(t\right)}{A\left(t\right)}=\frac{\dot{\phi}XG_{,X}\left(H-\dot{\phi}XG_{,X}\right)-\partial_{t}\left(H-\dot{\phi}XG_{,X}\right)}{XD}\,,\label{eq:speed_of_sound}
\end{equation}
where we assumed that the field $\phi$ is the only source of energy-momentum.
From this expression we can write 
\begin{equation}
\frac{1}{2}B\left(t\right)=\frac{\dot{\phi}XG_{,X}\left(H-\dot{\phi}XG_{,X}\right)-\partial_{t}\left(H-\dot{\phi}XG_{,X}\right)}{\left(H-\dot{\phi}XG_{,X}\right)^{2}}\text{ .}\label{eq:B_kgb}
\end{equation}
It is natural to introduce a quantity 
\begin{equation}
\gamma=H-\dot{\phi}XG_{,X}\text{ ,}\label{eq:gamma}
\end{equation}
in terms of which the expression for $B$ reads 
\begin{equation}
\frac{1}{2}B\left(t\right)=\frac{d}{dt}\gamma^{-1}+H\gamma^{-1}-1\text{ .}\label{eq:B_general}
\end{equation}
The expression for $B\left(t\right)$ was written in this elegant
form in \cite{Ijjas:2016tpn} for a particular choice (\ref{eq:KGB_identifiaction})
of functions $K$ and $G$. Before that this variable (\ref{eq:gamma})
was used in \cite{Libanov:2016kfc} and \cite{Quintin:2015rta}. The
vanishing $\gamma$ corresponds to the change of the branch in the
solution of the first Friedmann equation with respect to the Hubble
parameter (\ref{eq:Friedmann_with_branches}). In that case one cannot
express the perturbation of the lapse $\delta N$ from the momentum
constraint (\ref{eq:constraint_momentum}). Thus one has to use other
dynamical variables to describe the dynamics around this point. There
is an interesting discussion \cite{Battarra:2014tga,Quintin:2015rta,Ijjas:2017pei}
of gauge issues, choice of dynamical variables and slicing around
$\gamma=0$. This phenomenon is not special to \emph{kinetically braided}
theories. A spatially-flat Friedmann universe driven by a scalar field
with canonical kinetic term and a negative potential can evolve from
expansion to contraction, see e.g. \cite{Felder:2002jk}. At the turning
point $\gamma=H=0$, and one cannot exclude $\delta N$ from the action
for perturbations. 

For the Lagrangian given by (\ref{eq:KGB_identifiaction}) one obtains
\begin{equation}
\gamma\left(t\right)=H-\dot{\phi}X\,,\label{eq:Gamma_IS_Lagrangian}
\end{equation}
so that (\ref{eq:B_kgb}) (or (\ref{eq:B_general})) yields
\begin{equation}
B=\frac{2X\left(k+2qX+4H\dot{\phi}+2\ddot{\phi}-2X^{2}\right)}{\left(H-\dot{\phi}X\right)^{2}}\,,\label{eq:B_IS}
\end{equation}
while (\ref{eq:Our_Normalization}) reads 
\begin{equation}
A=\frac{2X\left(k+6qX+6\dot{\phi}H+6X^{2}\right)}{\left(H-\dot{\phi}X\right)^{2}}\,.\label{eq:Normalization}
\end{equation}

It is also useful to rewrite the action (\ref{eq:quadratic_action_scalars})
in terms of the canonically-normalized Mukhanov-Sasaki variable
\begin{equation}
v=z\mathcal{R}\text{ ,}\label{eq:MS_variable}
\end{equation}
where we denoted 
\[
z=a\sqrt{A}=a\sqrt{\frac{2XD}{\gamma^{2}}}\text{ .}\text{ }
\]
Then the action reads 
\begin{equation}
S_{c}=\frac{1}{2}\int d\tau\,d^{3}x\,\left(v'^{2}-c_{s}^{2}\left(\partial_{i}v\right)^{2}+\frac{z''}{z}v^{2}\right)\,,\label{eq:canonical_MS}
\end{equation}
where the prime denotes a derivative with respect to conformal time
$\tau$, defined through $d\tau=dt/a$ . 

It is worth noting that vanishing $D$ generically corresponds to
infinities of the square of the sound speed, (\ref{eq:speed_of_sound})
and to an infinitely strong coupling between canonically normalized
perturbations $v$. 

\section{Conformal Transformations, Gauges and Acoustic Geometry\label{sec:Conformal-Transformations,-Gauge} }

As this paper discusses superluminality, it is interesting to look
at the effective (or acoustic) metric where the scalar perturbations
propagate. It is possible to write the action (\ref{eq:quadratic_action_scalars})
in an elegant way 
\begin{equation}
S_{c}=\frac{1}{2}\int d^{4}x\sqrt{-\mathcal{G}}\text{ }\mathcal{G}^{\mu\nu}\partial_{\mu}\mathcal{R}\partial_{\nu}\mathcal{R}\,,\label{eq:Covariant_action_R}
\end{equation}
where the \emph{covariant} \emph{acoustic} metric $\mathcal{G}_{\mu\nu}^{-1}$
for curvature perturbations is 
\begin{equation}
dL^{2}=\mathcal{G}_{\mu\nu}^{-1}dx^{\mu}dx^{\nu}=z^{2}c_{s}\left(c_{s}^{2}d\tau^{2}-d\mathbf{x}^{2}\right)\text{ ,}\label{eq:acoustic_metric}
\end{equation}
the \emph{contravariant} metric is inverse to it and the notation
is standard. This metric $\mathcal{G}_{\mu\nu}^{-1}$ is singular
for $\gamma=0$, as $z\rightarrow\infty$. It seems that classically
this singularity is not a problem \cite{Battarra:2014tga,Quintin:2015rta,Ijjas:2017pei}.
Though it is interesting to understand the quantum mechanical consequences
of this singular behavior. As this study goes beyond the scope of
this paper, we leave this for a future work. 

The transformation to the canonical Mukhanov-Sasaki variable can be
considered as a \emph{conformal} transformation of the acoustic metric
\[
\mathcal{G}_{\mu\nu}^{-1}\rightarrow\ell_{\mu\nu}^{-1}=z^{-2}\mathcal{G}_{\mu\nu}^{-1}\,,\qquad\mathcal{R}\rightarrow v=z\mathcal{R}\text{ .}\text{ }
\]
The gauge part of the metric follows the propagation of the fluctuations
of the field $\delta\phi$. In the unitary gauge this is not obvious
as $\delta\phi=0$. However, one can perform a gauge transformation,
see e.g. page 293 \cite{Mukhanov:2005sc}, so that $\widetilde{\delta\phi}=-\phi'\xi^{0}$
and $\tilde{\psi}=\psi+\xi^{0}a'/a=-\mathcal{R}+\xi^{0}a'/a$. For
example for $\xi^{0}=\mathcal{R}a/a'$, the spatial metric becomes
unperturbed, $\delta g_{ik}=0$, and so in this ``spatially flat''
gauge 
\[
\left.\delta\phi\right|_{flat}=-\frac{\dot{\phi}}{H}\mathcal{R}\text{ .}
\]
Using analogy between $\gamma$ in \emph{braided} models and $H$
in k-essence we can introduce the ``$\gamma$-gauge'' 
\begin{equation}
\xi^{0}=\frac{\mathcal{R}}{a\gamma}\text{ ,}\label{eq:gamma-gauge}
\end{equation}
which yields
\[
\tilde{\varphi}=\varphi-\frac{1}{a}\left(a\xi^{0}\right)'=\delta N-\frac{1}{a}\left(a\xi^{0}\right)'=\mathcal{R}\frac{\dot{\gamma}}{\gamma^{2}}\text{ ,}
\]
where we used the constraint (\ref{eq:constraint_momentum}), and
\begin{equation}
\left.\delta\phi\right|_{\gamma}=-\frac{\dot{\phi}}{\gamma}\mathcal{R}\text{ .}\label{eq:scalar_fluctuations_gamma}
\end{equation}

Using $\left.\delta\phi\right|_{\gamma}$ the action (\ref{eq:quadratic_action_scalars})
reads 
\begin{equation}
S_{c}=\frac{1}{2}\int d^{4}x\sqrt{-\mathcal{G}}\text{ }\mathcal{G}^{\mu\nu}\text{ }\partial_{\mu}\mathcal{R}\text{ }\partial_{\nu}\mathcal{R}=\frac{1}{2}\int d^{4}x\sqrt{-C}\text{ }C^{\mu\nu}\text{ }\partial_{\mu}\mathcal{\left.\delta\phi\right|_{\gamma}}\text{ }\partial_{\nu}\mathcal{\left.\delta\phi\right|_{\gamma}}+...\,\,,\label{eq:Actions_Comparison}
\end{equation}
where the ellipsis stands for a ``mass-like'' term without derivatives
of $\mathcal{\left.\delta\phi\right|_{\gamma}}$. The change of variable
$\mathcal{R}\longleftrightarrow\left.\delta\phi\right|_{\gamma}$
can be understood as another \emph{conformal} transformation 

\[
\mathcal{G}_{\mu\nu}^{-1}\rightarrow C_{\mu\nu}^{-1}=\omega^{2}\mathcal{G}_{\mu\nu}^{-1}\,,\qquad\mathcal{R}\rightarrow\left.\delta\phi\right|_{\gamma}=-\omega^{-1}\mathcal{R}\text{ ,}\text{ }
\]
where $\omega=\gamma/\dot{\phi}$ so that 
\begin{equation}
dC^{2}=C_{\mu\nu}^{-1}dx^{\mu}dx^{\nu}=Dc_{s}a^{2}\left(c_{s}^{2}d\tau^{2}-d\mathbf{x}^{2}\right)\text{ .}\label{eq:acoustic_metric_field}
\end{equation}
 The acoustic metric $C_{\mu\nu}^{-1}$ differs from the metric given
by the formula (3.15), \cite{Deffayet:2010qz} by the normalization
$D^{2}c_{s}^{3}$. This conformal factor is not important for the
propagation of the high frequency perturbations, and related stability
studies, but is needed for a proper normalization of the action. This
transformation provides a short explanation for the so-called ``DPSV
trick'' discussed in \cite{Kolevatov:2017dze}. It is instructive
to compare this acoustic metric with the one obtained for k-essence
and cosmological perturbations \cite{Babichev:2007dw}, Appendix C.
There it was demonstrated that $\left.\delta\phi\right|_{flat}$ propagate
in the acoustic metric (\ref{eq:acoustic_metric_field}) with $D=\varepsilon_{,X}=K_{,X}+2XK_{,XX}$.
There is a continuity in $G,_{X}$ between ``$\gamma$-gauge'' and
``flat gauge''. 

The acoustic metric derived in \cite{Deffayet:2010qz} is generic
and can be used to investigate the speed of propagation of fluctuations,
gradient (in)stabilities and possible appearance of ghosts also around
general inhomogeneous and anisotropic backgrounds. In particular,
this check enables one to exclude wormholes \cite{Rubakov:2015gza,Rubakov:2016zah}
and static semiclosed worlds \cite{Evseev:2016ppw}. The advantage
of the acoustic metric is that it can be used for stability checks
for high frequency perturbations without deriving the action for perturbations.
The latter can be especially complicated in the presence of external
matter. 

Finally it is worth mentioning that one can express perturbations
through gauge-invariant variables which coincide with the conformal
Newtonian gauge (notation as in \cite{Mukhanov:2005sc})
\[
-\mathcal{R}=\Psi+\frac{H}{\dot{\phi}}\overline{\delta\phi}\text{ ,}
\]
and respectively 
\[
\left.\delta\phi\right|_{\gamma}=\frac{H}{\gamma}\text{ }\overline{\delta\phi}+\frac{\dot{\phi}}{\gamma}\Psi\text{ .}
\]

\section{Inverse method, finding the theory\label{sec:Inverse-method} }

\selectlanguage{british}%
\begin{figure}[t]
\selectlanguage{english}%
\subfloat[\foreignlanguage{american}{\label{fig:AB} Coefficients $A\left(t\right)$ and $B\left(t\right)$
on solution (\ref{eq:phi_solution}),}]{

\includegraphics[width=0.45\columnwidth]{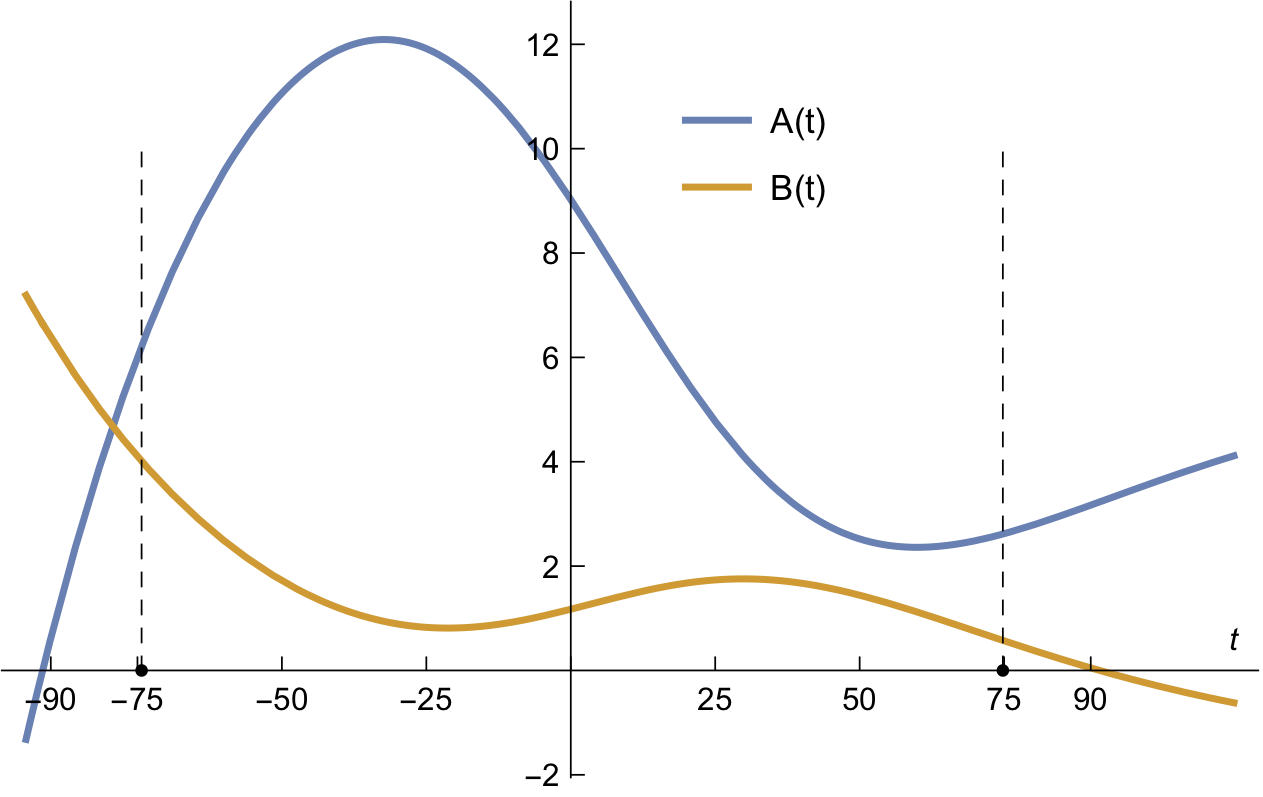}}\hfill{}\subfloat[\foreignlanguage{american}{\label{fig:Speed}$c_{s}^{2}\left(t\right)$ on the solution (\ref{eq:phi_solution})}]{

\includegraphics[width=0.45\columnwidth]{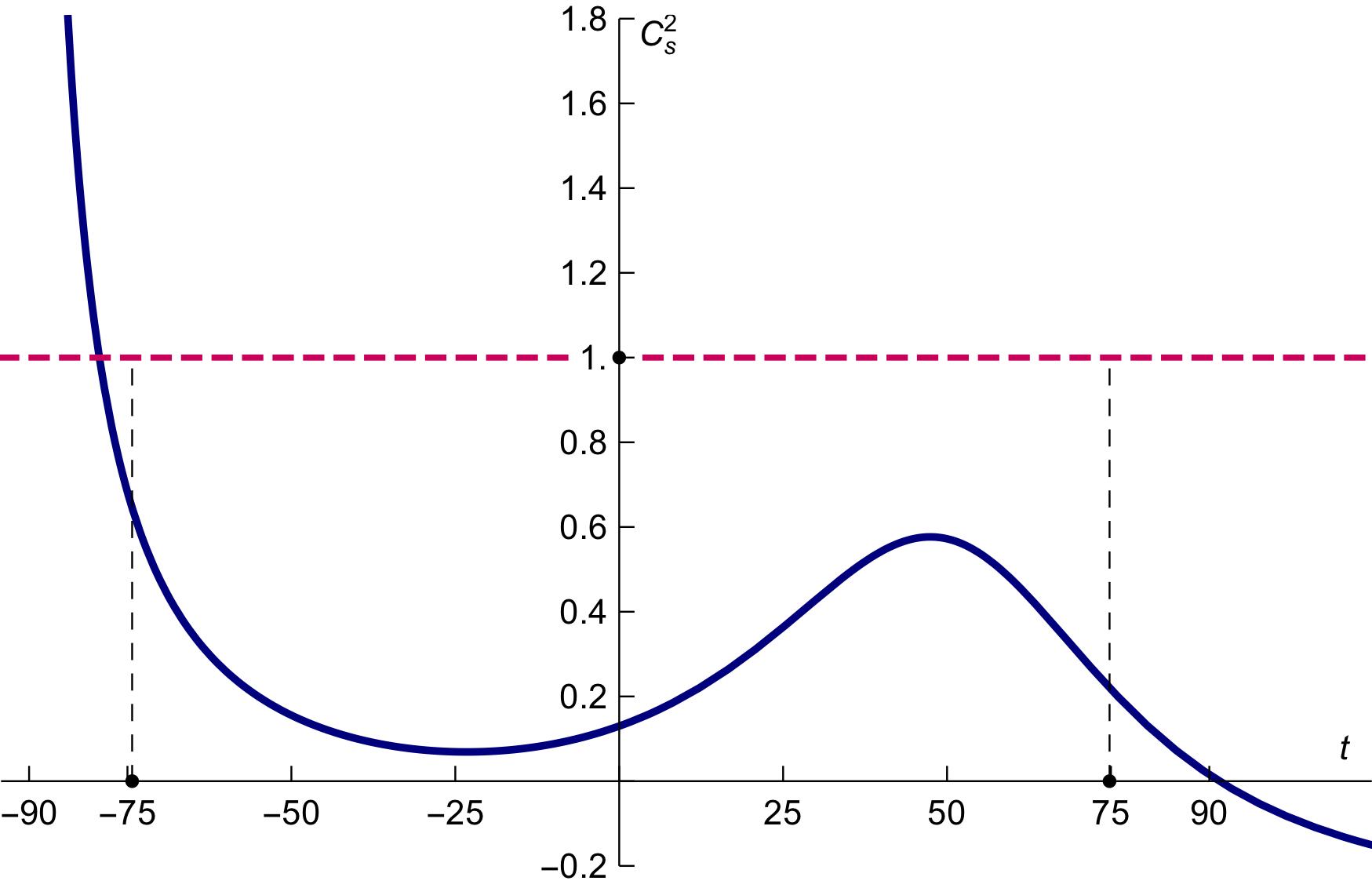}}\hfill{}

\selectlanguage{british}%
\caption{\foreignlanguage{american}{\label{IS_solution_perturbations_coeff} The evolution of the coefficients
in the quadratic action (\ref{eq:quadratic_action_scalars}) for the
scalar perturbations is shown here for the IS solution for the choice
of parameters (\ref{eq:First_choice}). All quantities are in the
Planck units. The dashed black vertical lines correspond to $t_{-}\simeq-74$
where the Phantom stage with NEC violation starts and to $t_{+}\simeq75$
where the NEC gets restored. Our right figure corresponds to Fig.
1 from \cite{Ijjas:2016tpn} just with a slightly extended time range.
Clearly, less than $10$ $t_{pl}$ before the beginning of the Phantom
stage the sound speed is \emph{superluminal}. On top of that, just
15 $t_{pl}$ after NEC is restored and the bouncing phase is finished
the system enters into an elliptic regime / regime with a gradient
instability: where $B<0$ and respectively $c_{s}^{2}<0$. When the
system approaches the regime, where $c_{s}^{2}=0$, the quantum fluctuations
naively \emph{diverge} and the system becomes strongly coupled. In
this case the semiclassical equations completely lose predictability.
On the other hand, just some 15 $t_{pl}$ before the beginning of
the Phantom stage the coefficient $A\left(t\right)$ vanishes and
the sound speed blows up. In order to enter the unmodified Phantom
bouncing stage and leave it without either starting or ending in these
strongly-coupled regimes one has to modify dynamics on time-scales
of $10$ $t_{pl}$ what is a clear challenge for the scenario. \protect \\
}}
\end{figure}

\selectlanguage{american}%
\begin{figure}[tb]
\begin{centering}
\includegraphics[width=0.5\paperwidth]{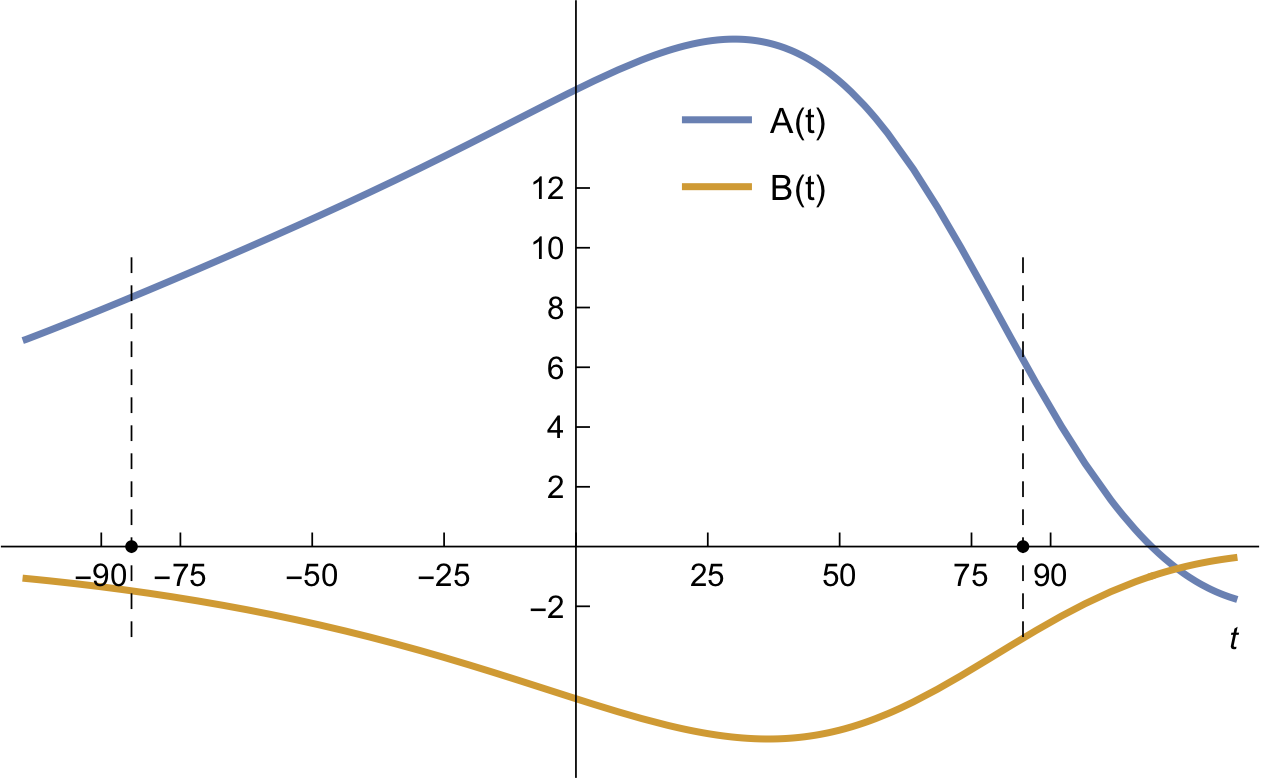}
\par\end{centering}
\caption{\foreignlanguage{british}{\label{fig:AB_Second_Choice} \foreignlanguage{american}{The evolution
of the coefficients in the quadratic action (\ref{eq:quadratic_action_scalars})
for the scalar perturbations is shown here for the IS solution for
the second choice of parameters (\ref{eq:Second_choice}). All quantities
are in the Planck units. The dashed black vertical lines correspond
to $t_{-}\simeq-84.3$ where the Phantom stage with NEC violation
starts and to $t_{+}\simeq84.8$ where the NEC gets restored. All
quantities are obtained using the same code as in the previous case.
The caption of the Fig. 3 of \cite{Ijjas:2016tpn} claims ``all fundamental
physical quantities including $H(t)$ and $c_{s}^{2}$ remain finite
and positive'' for this choice of parameters. On this plot one can
clearly see that $A\left(t\right)>0$, whereas $B\left(t\right)<0$
during the whole Phantom stage so that $c_{s}^{2}=B/A<0$. It seems
that there is a typo somewhere among the values of parameters (\ref{eq:Second_choice}).
}\protect \\
}}
\end{figure}

The authors of IS-bounce postulated a fairly simple time-dependence
of the Hubble parameter 
\begin{equation}
H\left(t\right)=H_{0}\,t\,\exp\left(-F\left(t-t_{*}\right)^{2}\right)\,,\label{eq:Hubble_IS}
\end{equation}
where $H_{0}$, $F$ and $t_{*}$ are constants. They proposed an
``inverse method'' to find free functions $k\left(\phi\right)$
and $q\left(\phi\right)$ in (\ref{eq:IS_Lagrangian}) which can realize
this cosmological evolution.  The NEC is violated between $t_{-}$
and $t_{+}$ where 
\begin{equation}
t_{\pm}=\frac{t_{*}\pm\sqrt{t_{*}^{2}+2F^{-1}}}{2}\text{ .}\label{eq:timespan_NEC_broken}
\end{equation}

The bounce occurs at $t=0$. For the bounce one has to start form
$H_{-}$ branch of the solutions of the first Friedmann equation (\ref{eq:Friedmann_with_branches}).
The key observation of the ``inverse method'' proposed in \cite{Ijjas:2016tpn}
is that one can also independently postulate $\gamma\left(t\right)$
in (\ref{eq:B_general}). The IS-bounce postulates 
\begin{equation}
\gamma=\gamma_{0}\exp\left(3\theta t\right)+H\left(t\right)\,,\label{eq:Gamma _IS}
\end{equation}
where $\gamma_{0}$ and $\theta$ are additional constants with respect
to already introduced $H_{0}$, $F$ and $t_{*}$. From (\ref{eq:Gamma_IS_Lagrangian})
one can obtain 
\[
\phi_{IS}\left(t\right)=\phi_{0}+\int_{t_{0}}^{t}dt'\left[2\left(H\left(t'\right)-\gamma\left(t'\right)\right)\right]^{1/3}\,,
\]
where $\phi_{IS}\left(t_{0}\right)=\phi_{0}$. It is convenient to
choose this initial value as $\phi_{0}=\left(-2\gamma_{0}/\theta\right)^{1/3}\exp\left(3t_{0}\right),$
so that the particular solution postulated in IS-bounce is 
\begin{equation}
\phi_{IS}\left(t\right)=\phi_{\star}\exp\left(\theta t\right)\,,\label{eq:phi_solution}
\end{equation}
where the field value at the bounce $\phi_{\star}$ is given by
\[
\phi_{\star}=\left(\frac{-2\gamma_{0}}{\theta^{3}}\right)^{1/3}\,.
\]
Then cosmological time is expressed on the IS-bounce as 
\begin{equation}
t=\frac{1}{\theta}\text{log}\left(\phi_{IS}/\phi_{\star}\right)\,.\label{eq:time_IS_solution}
\end{equation}
The field values $\phi_{1}$ and $\phi_{2}$ corresponding to the
beginning of the NEC violation and its restoration are given by 
\begin{align}
\phi_{1}=\phi_{IS}\left(t_{-}\right)\text{ ,}\quad & \phi_{2}=\phi_{IS}\left(t_{+}\right)\text{ .}\label{eq:field_range}
\end{align}
One has to rely on the reconstruction at least during the NEC violation,
i.e. in this field range $\phi_{1}<\phi<\phi_{2}$. Using the substitutions
(\ref{eq:Hubble_IS}) and (\ref{eq:Gamma _IS}) one obtains the functions
$k\left(\phi\right)$ and $q\left(\phi\right)$ as functions of time
on the particular solution (\ref{eq:phi_solution})
\begin{equation}
k\left(t\right)=-\frac{2\left(2\dot{H}+3H^{2}+\dot{\gamma}+3H\gamma\right)}{\left(2\left(H-\gamma\right)\right)^{2/3}}\,,\label{eq:k(t)}
\end{equation}
and 
\begin{equation}
q\left(t\right)=\frac{4\left(2\dot{H}+\dot{\gamma}+9H\gamma\right)}{3\left(2\left(H-\gamma\right)\right)^{4/3}}\,.\label{eq:q(t)}
\end{equation}
It is convenient to introduce functions 
\begin{equation}
W\left(\phi\right)=\exp\left[\frac{F}{\theta^{2}}\left(\log\left(\frac{\phi}{\phi_{\star}}\right)-\theta t_{*}\right){}^{2}\right]\,,\label{eq:W_IS}
\end{equation}
and 
\begin{equation}
\Omega\left(\phi\right)=W\left(\phi\right)\theta^{3}\left(\theta\phi\right)^{3}+H_{0}\left[\log\left(\frac{\phi}{\phi_{\star}}\right)\left(4F\left[\log\left(\frac{\phi}{\phi_{\star}}\right)-\theta t_{*}\right]+2\theta\left(\theta\phi\right)^{3}\right)-2\theta^{2}\right]\,,\label{eq:Omega_IS}
\end{equation}
in terms of which the defining functions are 
\begin{equation}
k(\phi)=-\frac{12H_{0}^{2}\log^{2}\left(\phi/\phi_{\star}\right)-3W\left(\phi\right)\left[\Omega\left(\phi\right)-H_{0}\theta\left(\theta\phi\right)^{3}\log\left(\phi/\phi_{\star}\right)\right]}{W^{2}\left(\phi\right)\theta^{2}\left(\theta\phi\right){}^{2}}\,,\label{eq:k(phi)}
\end{equation}
and 
\begin{equation}
q\left(\phi\right)=\frac{12H_{0}^{2}\log^{2}\left(\phi/\phi_{\star}\right)-2W\left(\phi\right)\left[\Omega\left(\phi\right)+H_{0}\theta\left(\theta\phi\right)^{3}\log\left(\phi/\phi_{\star}\right)\right]}{W^{2}\left(\phi\right)\theta^{2}\left(\theta\phi\right){}^{4}}\,.\label{eq:q(phi)}
\end{equation}
These expressions defining the theory, which should be related to
the very origin of the universe, neither look well-motivated nor natural
from any point of view. Neither these functions can be stable with
respect to the quantum corrections. This is the price for the chosen
simple exact solution (\ref{eq:Hubble_IS}), (\ref{eq:phi_solution}).
We are left with the following five free parameters $H_{0}$, $\theta$,
$F$, $\gamma_{0}$, $t_{*}$ which specify the Lagrangian. The authors
of \cite{Ijjas:2016tpn} have chosen them below their Fig. 1 as 
\begin{equation}
H_{0}=3\times10^{-5}\;,\theta=0.0046\;,F=9\times10^{-5}\;,\gamma_{0}=-0.0044\;,t_{*}=0.5\text{ ,}\label{eq:First_choice}
\end{equation}
all in reduced Planck units. The role of this unphysically tiny $t_{*}$
remained an open question for us. For this choice of parameters the
field value at the bounce is $\phi_{\star}\simeq44.88$ and $t_{-}\simeq-74.286$
while $t_{+}\simeq74.786$. Now we can plot the coefficients $A\left(t\right)$
and $B\left(t\right)$ and the sound speed $c_{s}^{2}$ on the trajectory
(\ref{eq:phi_solution}), see Fig. (\ref{fig:AB}) and Fig. (\ref{fig:Speed})
respectively. On these plots we clearly see that less than $10$ $t_{pl}$
before the beginning of the Phantom stage the sound speed is \emph{superluminal}.
Moreover, just 15 $t_{pl}$ after NEC is restored and the bouncing
phase is finished the system enters into elliptic regime / regime
with the gradient instability. When the system approaches the regime,
where $c_{s}^{2}=0$ the quantum perturbations \emph{diverge} and
the system becomes strongly coupled. In that case the semiclassical
equations completely lose predictability. On the other hand just some
15 $t_{pl}$ before the beginning of the Phantom stage the coefficient
$A\left(t\right)$ vanishes and the sound speed becomes not only just
superluminal, but simply \emph{divergent}. Both divergent and vanishing
sound speeds correspond to infinitely strongly coupled fluctuations.
Indeed, in weakly coupled theories on short length scales $\ell$,
where $\left(kc_{s}\right)^{2}\gg\left|z''/z\right|$, one can use
the uncertainty relation \cite{Vikman:2012bx,Easson:2016klq} to find
\begin{equation}
\delta v_{\ell}\cdot\delta v'_{\ell}\simeq\hbar\,\ell^{-3}\text{ .}\label{eq:uncertainty}
\end{equation}

Further estimating $\delta v'_{\ell}\simeq\omega_{\ell}\delta v_{\ell}\simeq c_{s}\ell^{-1}\delta v_{\ell}$
we obtain 
\begin{align}
\delta v_{\ell}\simeq\ell^{-1}\sqrt{\hbar/c_{s}}\text{ ,} & \qquad\delta v'_{\ell}\simeq\ell^{-2}\sqrt{\hbar c_{s}}\text{ .}\label{eq:Q_fluctuations}
\end{align}
Hence larger values of $c_{s}$ correspond to larger velocity fluctuation
$\delta v'_{\ell}$ on all short scales. Whereas vanishingly small
$c_{s}$ corresponds to a huge $\delta v_{\ell}$. In these non-canonical
theories both the field and the canonical momentum do enter the interaction
vertices. Clearly very large (divergent) quantum fluctuations is a
pathology. The only way to avoid these estimations is to assume that
the theory is strongly coupled, so that the uncertainty relation is
not saturated and that the fluctuation of momentum $\delta v'_{\ell}$
is not related to the fluctuation of the field $\delta v_{\ell}$
as it is in the quantum oscillator case. But then the theory is clearly
strongly coupled in the quantum mechanical sense. In order to enter
the unmodified Phantom bouncing stage and leave it without either
starting or ending in these strongly-coupled regimes one has to modify
dynamics on time-scales of $10$ $t_{pl}$ which is a clear challenge
for the scenario.

The second example of IS was the set of parameters chosen below their
Fig. 3 as 
\begin{equation}
H_{0}=3\times10^{-5}\;,\theta=4.6\times10^{-6}\;,F=7\times10^{-5}\;,\gamma_{0}=-0.0044\;,t_{*}=0.5\text{ .}\label{eq:Second_choice}
\end{equation}
For the corresponding times we have $t_{-}\simeq-84.266$ while $t_{+}\simeq84.766$.
Clearly the caption of the Fig. 3 of \cite{Ijjas:2016tpn} claims
``all fundamental physical quantities including $H(t)$ and $c_{s}^{2}$
remain finite and positive'' for this choice of parameters. However,
we found that the sound speed is actually imaginary throughout the
NEC-violating stage, see Fig. (\ref{fig:AB_Second_Choice}) for the
coefficients $A\left(t\right)$ and $B\left(t\right)$. Clearly $A\left(t\right)>0$
whereas $B\left(t\right)<0$ during whole phantom stage. Hence the
sound speed $c_{s}^{2}=B/A<0$. It seems that there is somewhere a
typo in these values of parameters (\ref{eq:Second_choice}). 

\section{Phase Space}

\selectlanguage{british}%
\begin{figure}[t]
\selectlanguage{english}%
\subfloat{\includegraphics[width=0.45\columnwidth]{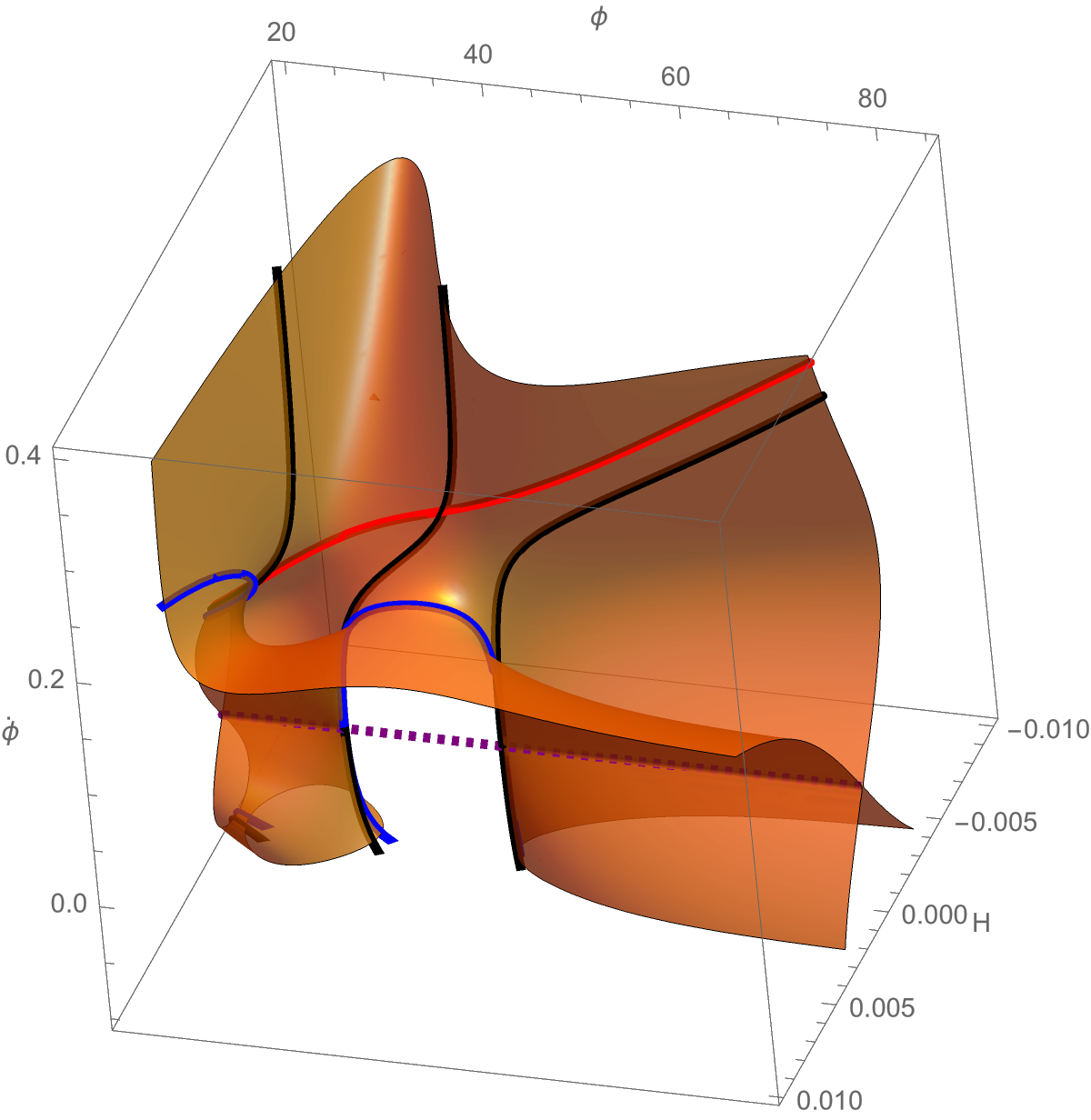}}\hfill{}\subfloat{\includegraphics[width=0.45\columnwidth]{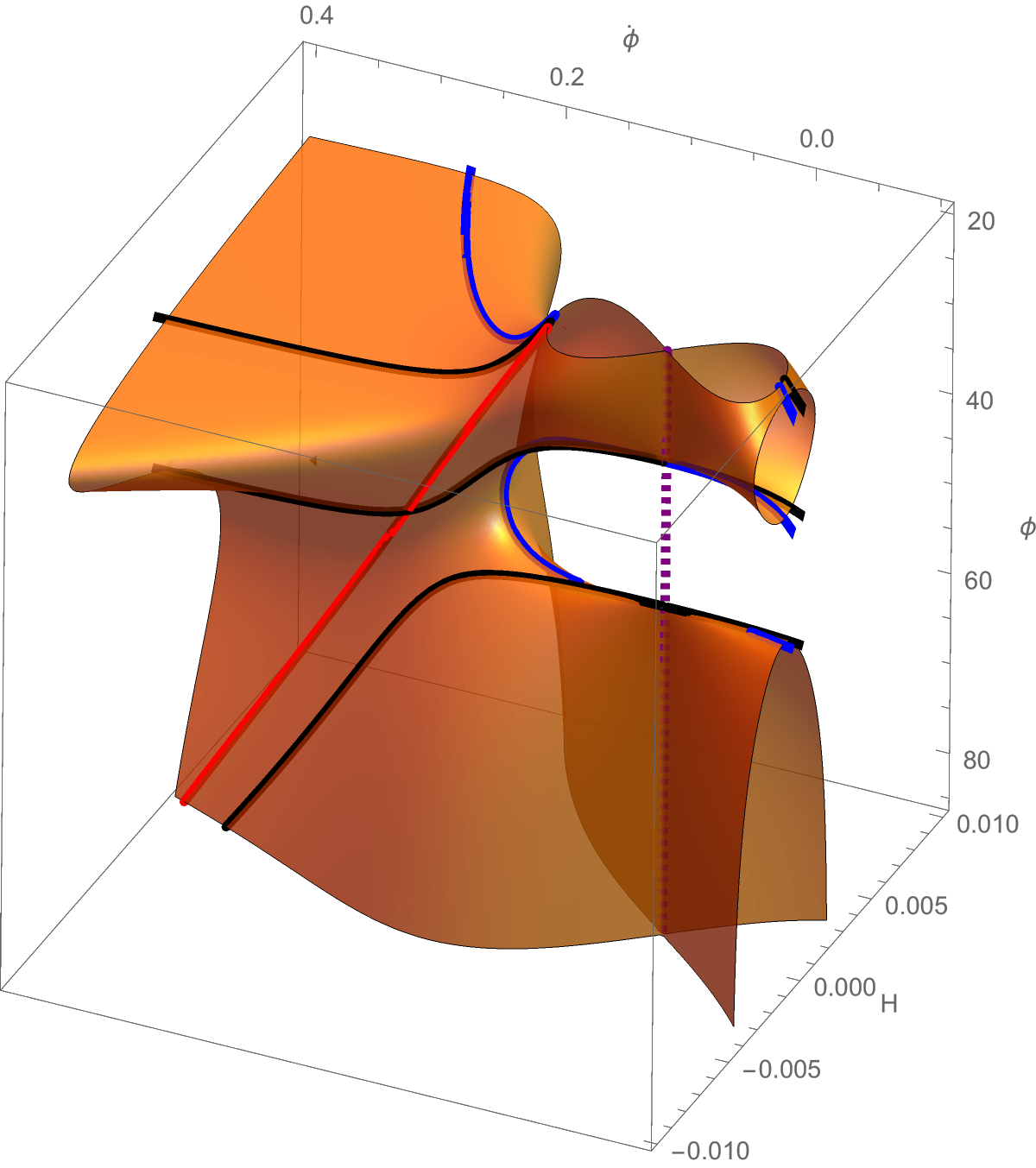}}\hfill{}

\selectlanguage{british}%
\caption{\foreignlanguage{american}{\label{fig:Surface} Here we plot different side views of the phase
space hypersurface given by the constraint - the but first Friedmann
equation (\ref{eq:First_Friedmann}) for the system defined by (\ref{eq:IS_Lagrangian})
and (\ref{eq:k(phi)}), (\ref{eq:q(phi)}). The parameters correspond
to the choice of \cite{Ijjas:2016tpn} below their Fig. 1, see (\ref{eq:First_choice}).
The red curve is the-IS bounce trajectory. The black curves correspond
to $H=0$ while the blue curves represent $\gamma=0$. The purple
dashed lines represent $\dot{\phi}=0$. Each point on these lines
on the hypersurface of the constraint is a fixed point. Therefore
the self-crossing of the hypersurface does not cause any trouble. }}
\end{figure}

\selectlanguage{american}%
\begin{figure}[tb]
\begin{centering}
\includegraphics[width=0.75\paperwidth]{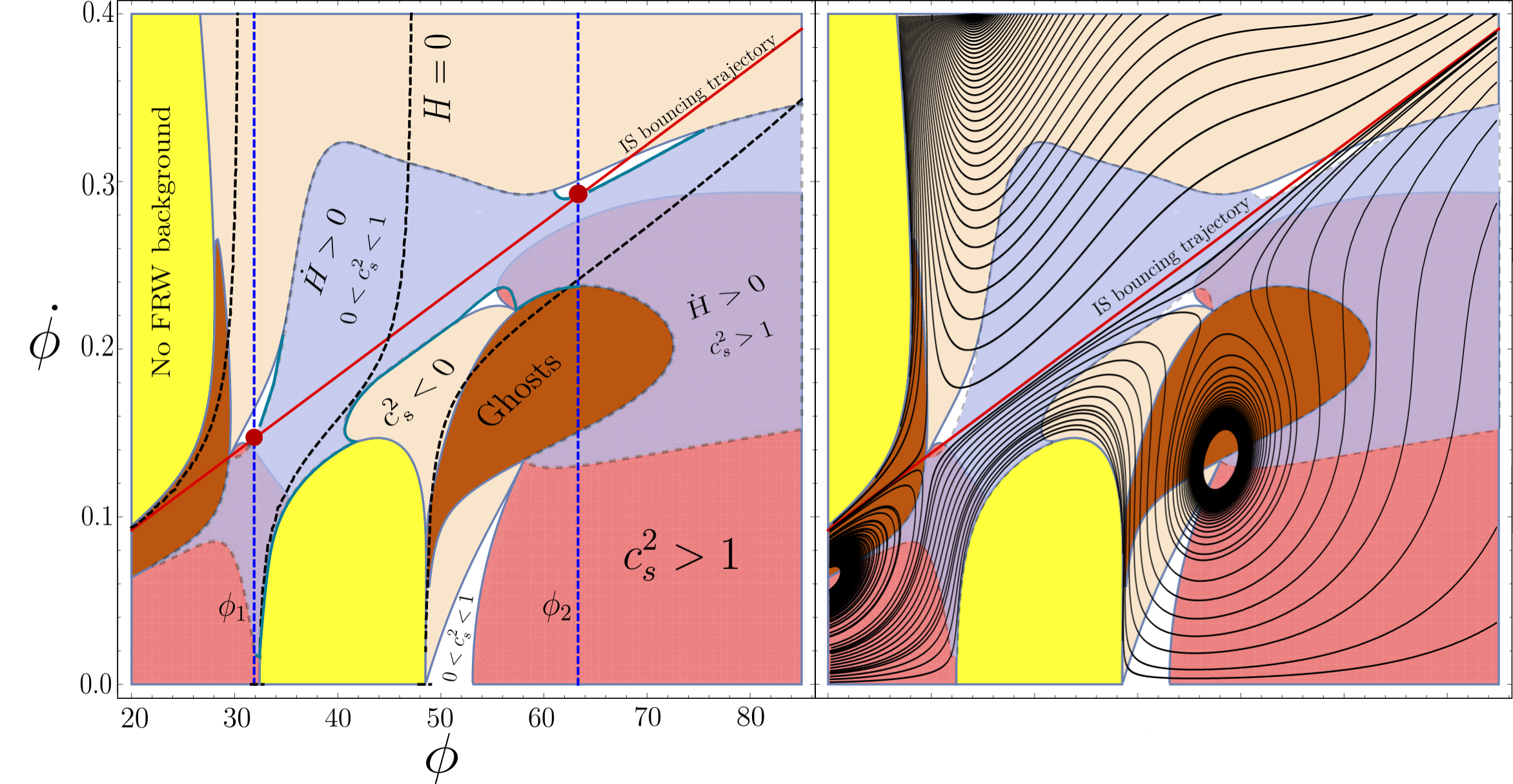}
\par\end{centering}
\caption{\foreignlanguage{british}{\label{fig:PhaseSpace} \foreignlanguage{american}{Here we plot the
phase space for the system defined by (\ref{eq:IS_Lagrangian}) and
(\ref{eq:k(phi)}), (\ref{eq:q(phi)}). We pick the $H_{-}$ branch
in (\ref{eq:Friedmann_with_branches}). The parameters correspond
to the choice of \cite{Ijjas:2016tpn} below their Fig. 1, see (\ref{eq:First_choice}).
This plot is a projection of the hypersurface from the Fig. (\ref{fig:Surface}).
The red line corresponds to the \emph{IS-Bounce}. This bouncing trajectory
is a separatrix which goes from a saddle point, see the plot on the
right. In the \emph{yellow} regions, the condition (\ref{eq:Friedmann_with_branches})
or (\ref{eq:allowed_space}) is not fulfilled, so that there is no
spatially-flat FRW geometry there. The phase space continues to the
other branch of the Friedmann equation (\ref{eq:Friedmann_with_branches})
through the borders of these regions where $\gamma=0$. In the \emph{light
brown / almond }regions $c_{s}^{2}<0$, and the system has a gradient
instability. The borders of these regions correspond to $c_{s}^{2}=0$
which causes an infinitely strong coupling of curvature perturbations.
The \emph{burned orange }/ \emph{dark brown} regions have ghosts,
$D<0$ there, see (\ref{eq:D_for_ghosts}) and (\ref{eq:D_IS-1}).
The boundaries of these regions have $D=0$ which implies an infinite
pressure (\ref{eq:pressure}) and correspondingly an infinite curvature.
These boundaries are also singularities of the background equations
of motion (\ref{eq:EoM_phi}). In the \emph{Congo pink / coral} regions
the sound speed is superluminal $c_{s}^{2}>1$, but the NEC holds.
Light blue /\emph{ lavender} regions correspond to the NEC violation
without superluminality and free of ghosts and gradient instabilities.
Purple / \emph{blue bell} regions have the NEC violation and superluminality,
but are free of ghosts and gradient instabilities. These superluminal
regions are located only slightly below the red \emph{IS-Bounce} trajectory.
Finally, four small \emph{white} regions are rather boring as they
are free of ghosts, gradient instabilities, superluminality and violation
of the NEC. The \emph{IS-Bounce} crosses two of these white regions.
On the red bouncing trajectory (\ref{eq:phi_solution}) the NEC is
broken between $\phi_{1}$ and $\phi_{2}$, (\ref{eq:field_range}).
One has to rely on the reconstruction of the Lagrangian functions
(\ref{eq:k(phi)}) and (\ref{eq:q(phi)}) \emph{at least} between
these two field values, between \emph{blue dashed} lines. The \emph{black
dashed} curves correspond to $H=0$. There are many trajectories above
and below the red trajectory and on the right of the $\phi_{2}$ which
bounce and evolve through the \emph{black dashed} curves. All trajectories
start or end (or both) on a singularity or infinitely-strong coupling
curves. }\protect \\
}}
\end{figure}

The reconstruction of the Lagrangian through the identification of
the functions (\ref{eq:k(t)}) and (\ref{eq:q(t)}) allows us to study
the properties of other cosmological solutions in the system under
consideration. A proper global analysis can follow the lines of \cite{Felder:2002jk}.
One chooses dynamical variables $\left(\phi,\dot{\phi},H\right)$
which evolution is given by the second Friedmann equation (\ref{eq:Second_Friedmann})
and the equation of motion for the scalar field (\ref{eq:EoM_phi})
written as a first order system. These dynamical variables are moving
on a hypersurface given by the constraint - first Friedmann equation
(\ref{eq:First_Friedmann}), see Fig. (\ref{fig:Surface}). In many
cases this hypersurface cannot be uniquely projected onto $\left(\phi,\dot{\phi}\right)$
plane, see discussion in \cite{Felder:2002jk} for a canonical scalar
field and \cite{Easson:2011zy,Easson:2013bda} for \emph{kinetically
braided} theories. In the latter it was found that sometimes one can
uniquely project the constraint hypersurface onto the space $\left(\dot{\phi},H\right)$.
To make a projection onto $\left(\phi,\dot{\phi}\right)$ plane, one
has to choose the branch in the solution of the first Friedmann equation
and substitute this branch $H_{\pm}$ into the reduced field equation
(\ref{eq:EoM_phi}). In that case, the dynamics are only present in
the region of the phase space where condition (\ref{eq:condition})
holds, i.e. where
\begin{equation}
\dot{\phi}^{4}+q\left(\phi\right)\dot{\phi}^{2}+\frac{2}{3}k\left(\phi\right)\geq0\text{ .}\label{eq:allowed_space}
\end{equation}
In the Fig. (\ref{fig:PhaseSpace}) we plot different regions in the
remaining phase space. We found other stable bouncing trajectories,
see the right plot on the Fig. (\ref{fig:PhaseSpace}). In this figure
one can see stable regions where superluminality is present. In parts
of these superluminal regions the NEC holds, while in others it is
violated. Moreover, there is a region of phase space where NEC is
broken but the sound speed is subluminal. This shows again that the
superluminality is not directly linked to stability, c.f. \cite{Dubovsky:2005xd}.
The link is rather subtle. In particular one can find the superluminality
just around the corner of the \emph{IS-bounce}, in the neighborhood
slightly below the trajectory.\emph{ }These superluminal stable regions
are well within the field range corresponding to the NEC violation
phase. Clearly a source or simple interaction can continuously deform
these states into the \emph{IS-bounce }trajectory\emph{. }Hence, these\emph{
}states belong to the same EFT. 

The origin of the trajectory is the ghosty region followed by a tiny
superluminal region where NEC holds. Then it is followed by a vanishingly
tiny white subluminal region where NEC holds. The \emph{IS-bounce}
trajectory leaves the ghosty region by going through the singularity
of the equation of motion $D=0$. This is also a pressure/ curvature
singularity. Thus this \emph{IS-bounce} trajectory is clearly demonstrating
the singular behavior similar to that crossing the phantom divide
in k-essence models linear in $X$ \cite{Vikman:2004dc}. It seems
that around the point on the boundary of the central ghosty region
the trajectories for a limiting cycle similarly to \cite{Easson:2016klq}.
This is, however, an illusion as they start and end on the singularity
approaching or leaving the boundary $D=0$ vertically. 

\acknowledgments The work of D.D. was funded by the Undergraduate
Research grant provided by the Natural Sciences and Engineering Research
Council of Canada. A.F and J.G. want to acknowledge the support of
the Discovery Grants by the Natural Sciences and Engineering Research
Council of Canada, J.G was partially funded by the Billy Jones scholarship
granted by the Department of Physics at Simon Fraser University and
by the Perimeter Institute for Theoretical Physics. Research at the
Perimeter Institute is supported by the Government of Canada through
the Department of Innovation, Science and Economic Development Canada.
The work of A.V. and S.R. was supported by the funds from the European
Regional Development Fund and the Czech Ministry of Education, Youth
and Sports (M\v{S}MT): Project CoGraDS - CZ.02.1.01/0.0/0.0/15\_003/0000437.
A.V. also acknowledges support from the J. E. Purkyn\v{e} Fellowship
of the Czech Academy of Sciences. It is a pleasure to thank Iggy Sawicki
for useful discussions. 

\bibliographystyle{utphys}
\addcontentsline{toc}{section}{\refname}\bibliography{IS_bounce}

\providecommand{\href}[2]{#2}\begingroup\raggedright\begin{thebibliography}{10}

\bibitem{Ijjas:2016tpn}
A.~Ijjas and P.~J. Steinhardt, ``{Classically stable nonsingular cosmological
  bounces},'' \href{http://dx.doi.org/10.1103/PhysRevLett.117.121304}{{\em
  Phys. Rev. Lett.} {\bfseries 117} no.~12, (2016) 121304},
\href{http://arxiv.org/abs/1606.08880}{{\ttfamily arXiv:1606.08880 [gr-qc]}}.

\bibitem{Deffayet:2010qz}
C.~Deffayet, O.~Pujolas, I.~Sawicki, and A.~Vikman, ``{Imperfect Dark Energy
  from Kinetic Gravity Braiding},''
  \href{http://dx.doi.org/10.1088/1475-7516/2010/10/026}{{\em JCAP} {\bfseries
  1010} (2010) 026},
\href{http://arxiv.org/abs/1008.0048}{{\ttfamily arXiv:1008.0048 [hep-th]}}.

\bibitem{Vikman:2004dc}
A.~Vikman, ``{Can dark energy evolve to the phantom?},''
  \href{http://dx.doi.org/10.1103/PhysRevD.71.023515}{{\em Phys. Rev.}
  {\bfseries D71} (2005) 023515},
\href{http://arxiv.org/abs/astro-ph/0407107}{{\ttfamily
  arXiv:astro-ph/0407107}}.

\bibitem{Hu:2004kh}
W.~Hu, ``{Crossing the phantom divide: Dark energy internal degrees of
  freedom},'' \href{http://dx.doi.org/10.1103/PhysRevD.71.047301}{{\em
  Phys.Rev.} {\bfseries D71} (2005) 047301},
  \href{http://arxiv.org/abs/astro-ph/0410680}{{\ttfamily
  arXiv:astro-ph/0410680 [astro-ph]}}.

\bibitem{Caldwell:2005ai}
R.~R. Caldwell and M.~Doran, ``{Dark-energy evolution across the
  cosmological-constant boundary},''
  \href{http://dx.doi.org/10.1103/PhysRevD.72.043527}{{\em Phys.Rev.}
  {\bfseries D72} (2005) 043527},
  \href{http://arxiv.org/abs/astro-ph/0501104}{{\ttfamily
  arXiv:astro-ph/0501104 [astro-ph]}}.

\bibitem{Xia:2007km}
J.-Q. Xia, Y.-F. Cai, T.-T. Qiu, G.-B. Zhao, and X.~Zhang, ``{Constraints on
  the Sound Speed of Dynamical Dark Energy},''
  \href{http://dx.doi.org/10.1142/S0218271808012784}{{\em Int. J. Mod. Phys.}
  {\bfseries D17} (2008) 1229--1243},
\href{http://arxiv.org/abs/astro-ph/0703202}{{\ttfamily
  arXiv:astro-ph/0703202}}.

\bibitem{Easson:2016klq}
D.~A. Easson and A.~Vikman, ``{The Phantom of the New Oscillatory Cosmological
  Phase},''
\href{http://arxiv.org/abs/1607.00996}{{\ttfamily arXiv:1607.00996 [gr-qc]}}.

\bibitem{deRham:2017aoj}
C.~de~Rham and S.~Melville, ``{Unitary null energy condition violation in P(X)
  cosmologies},'' \href{http://dx.doi.org/10.1103/PhysRevD.95.123523}{{\em
  Phys. Rev.} {\bfseries D95} no.~12, (2017) 123523},
\href{http://arxiv.org/abs/1703.00025}{{\ttfamily arXiv:1703.00025 [hep-th]}}.

\bibitem{Sawicki:2012pz}
I.~Sawicki and A.~Vikman, ``{Hidden Negative Energies in Strongly Accelerated
  Universes},'' \href{http://dx.doi.org/10.1103/PhysRevD.87.067301}{{\em
  Phys.Rev.} {\bfseries D87} (2013) 067301},
\href{http://arxiv.org/abs/1209.2961}{{\ttfamily arXiv:1209.2961
  [astro-ph.CO]}}.

\bibitem{Kobayashi:2010cm}
T.~Kobayashi, M.~Yamaguchi, and J.~Yokoyama, ``{G-inflation: Inflation driven
  by the Galileon field},''
  \href{http://dx.doi.org/10.1103/PhysRevLett.105.231302}{{\em Phys.Rev.Lett.}
  {\bfseries 105} (2010) 231302},
  \href{http://arxiv.org/abs/1008.0603}{{\ttfamily arXiv:1008.0603 [hep-th]}}.

\bibitem{Creminelli:2010ba}
P.~Creminelli, A.~Nicolis, and E.~Trincherini, ``{Galilean Genesis: An
  Alternative to inflation},''
  \href{http://dx.doi.org/10.1088/1475-7516/2010/11/021}{{\em JCAP} {\bfseries
  1011} (2010) 021}, \href{http://arxiv.org/abs/1007.0027}{{\ttfamily
  arXiv:1007.0027 [hep-th]}}.

\bibitem{Elder:2013gya}
B.~Elder, A.~Joyce, and J.~Khoury, ``{From Satisfying to Violating the Null
  Energy Condition},'' \href{http://dx.doi.org/10.1103/PhysRevD.89.044027}{{\em
  Phys. Rev.} {\bfseries D89} no.~4, (2014) 044027},
\href{http://arxiv.org/abs/1311.5889}{{\ttfamily arXiv:1311.5889 [hep-th]}}.

\bibitem{Rubakov:2014jja}
V.~A. Rubakov, ``{The Null Energy Condition and its violation},''
  \href{http://dx.doi.org/10.3367/UFNe.0184.201402b.0137}{{\em Phys. Usp.}
  {\bfseries 57} (2014) 128--142},
  \href{http://arxiv.org/abs/1401.4024}{{\ttfamily arXiv:1401.4024 [hep-th]}}.
[Usp. Fiz. Nauk184,no.2,137(2014)].

\bibitem{ArmendarizPicon:1999rj}
C.~Armendariz-Picon, T.~Damour, and V.~F. Mukhanov, ``{k-Inflation},''
  \href{http://dx.doi.org/10.1016/S0370-2693(99)00603-6}{{\em Phys. Lett.}
  {\bfseries B458} (1999) 209--218},
\href{http://arxiv.org/abs/hep-th/9904075}{{\ttfamily arXiv:hep-th/9904075}}.

\bibitem{ArmendarizPicon:2000ah}
C.~Armendariz-Picon, V.~F. Mukhanov, and P.~J. Steinhardt, ``{Essentials of
  k-essence},'' \href{http://dx.doi.org/10.1103/PhysRevD.63.103510}{{\em Phys.
  Rev.} {\bfseries D63} (2001) 103510},
\href{http://arxiv.org/abs/astro-ph/0006373}{{\ttfamily
  arXiv:astro-ph/0006373}}.

\bibitem{ArmendarizPicon:2000dh}
C.~Armendariz-Picon, V.~F. Mukhanov, and P.~J. Steinhardt, ``{A dynamical
  solution to the problem of a small cosmological constant and late-time cosmic
  acceleration},'' \href{http://dx.doi.org/10.1103/PhysRevLett.85.4438}{{\em
  Phys. Rev. Lett.} {\bfseries 85} (2000) 4438--4441},
\href{http://arxiv.org/abs/astro-ph/0004134}{{\ttfamily
  arXiv:astro-ph/0004134}}.

\bibitem{Garriga:1999vw}
J.~Garriga and V.~F. Mukhanov, ``{Perturbations in k-inflation},''
  \href{http://dx.doi.org/10.1016/S0370-2693(99)00602-4}{{\em Phys. Lett.}
  {\bfseries B458} (1999) 219--225},
\href{http://arxiv.org/abs/hep-th/9904176}{{\ttfamily arXiv:hep-th/9904176}}.

\bibitem{Luty:2003vm}
M.~A. Luty, M.~Porrati, and R.~Rattazzi, ``{Strong interactions and stability
  in the DGP model},'' {\em JHEP} {\bfseries 0309} (2003) 029,
\href{http://arxiv.org/abs/hep-th/0303116}{{\ttfamily arXiv:hep-th/0303116
  [hep-th]}}.

\bibitem{Nicolis:2004qq}
A.~Nicolis and R.~Rattazzi, ``{Classical and quantum consistency of the DGP
  model},'' \href{http://dx.doi.org/10.1088/1126-6708/2004/06/059}{{\em JHEP}
  {\bfseries 0406} (2004) 059},
\href{http://arxiv.org/abs/hep-th/0404159}{{\ttfamily arXiv:hep-th/0404159
  [hep-th]}}.

\bibitem{Nicolis:2008in}
A.~Nicolis, R.~Rattazzi, and E.~Trincherini, ``{The galileon as a local
  modification of gravity},''
  \href{http://dx.doi.org/10.1103/PhysRevD.79.064036}{{\em Phys. Rev.}
  {\bfseries D79} (2009) 064036},
\href{http://arxiv.org/abs/0811.2197}{{\ttfamily arXiv:0811.2197 [hep-th]}}.

\bibitem{Pujolas:2011he}
O.~Pujolas, I.~Sawicki, and A.~Vikman, ``{The Imperfect Fluid behind Kinetic
  Gravity Braiding},'' \href{http://dx.doi.org/10.1007/JHEP11(2011)156}{{\em
  JHEP} {\bfseries 1111} (2011) 156},
\href{http://arxiv.org/abs/1103.5360}{{\ttfamily arXiv:1103.5360 [hep-th]}}.

\bibitem{Deffayet:2011gz}
C.~Deffayet, X.~Gao, D.~Steer, and G.~Zahariade, ``{From k-essence to
  generalised Galileons},''
  \href{http://dx.doi.org/10.1103/PhysRevD.84.064039}{{\em Phys.Rev.}
  {\bfseries D84} (2011) 064039},
\href{http://arxiv.org/abs/1103.3260}{{\ttfamily arXiv:1103.3260 [hep-th]}}.

\bibitem{Kobayashi:2011nu}
T.~Kobayashi, M.~Yamaguchi, and J.~Yokoyama, ``{Generalized G-inflation:
  Inflation with the most general second-order field equations},''
  \href{http://dx.doi.org/10.1143/PTP.126.511}{{\em Prog.Theor.Phys.}
  {\bfseries 126} (2011) 511--529},
\href{http://arxiv.org/abs/1105.5723}{{\ttfamily arXiv:1105.5723 [hep-th]}}.

\bibitem{Horndeski:1974}
G.~W. Horndeski, ``Second-order scalar-tensor field equations in a
  four-dimensional space,'' \href{http://dx.doi.org/10.1007/BF01807638}{{\em
  Int. J. Theor. Phys.} {\bfseries 10} (1974) 363--384}.

\bibitem{Qiu:2011cy}
T.~Qiu, J.~Evslin, Y.-F. Cai, M.~Li, and X.~Zhang, ``{Bouncing Galileon
  Cosmologies},'' \href{http://dx.doi.org/10.1088/1475-7516/2011/10/036}{{\em
  JCAP} {\bfseries 1110} (2011) 036},
\href{http://arxiv.org/abs/1108.0593}{{\ttfamily arXiv:1108.0593 [hep-th]}}.

\bibitem{Easson:2011zy}
D.~A. Easson, I.~Sawicki, and A.~Vikman, ``{G-Bounce},''
  \href{http://dx.doi.org/10.1088/1475-7516/2011/11/021}{{\em JCAP} {\bfseries
  1111} (2011) 021},
\href{http://arxiv.org/abs/1109.1047}{{\ttfamily arXiv:1109.1047 [hep-th]}}.

\bibitem{Libanov:2016kfc}
M.~Libanov, S.~Mironov, and V.~Rubakov, ``{Generalized Galileons: instabilities
  of bouncing and Genesis cosmologies and modified Genesis},''
  \href{http://dx.doi.org/10.1088/1475-7516/2016/08/037}{{\em JCAP} {\bfseries
  1608} no.~08, (2016) 037},
\href{http://arxiv.org/abs/1605.05992}{{\ttfamily arXiv:1605.05992 [hep-th]}}.

\bibitem{Kobayashi:2016xpl}
T.~Kobayashi, ``{Generic instabilities of nonsingular cosmologies in Horndeski
  theory: A no-go theorem},''
  \href{http://dx.doi.org/10.1103/PhysRevD.94.043511}{{\em Phys. Rev.}
  {\bfseries D94} no.~4, (2016) 043511},
\href{http://arxiv.org/abs/1606.05831}{{\ttfamily arXiv:1606.05831 [hep-th]}}.

\bibitem{Kolevatov:2016ppi}
R.~Kolevatov and S.~Mironov, ``{Cosmological bounces and Lorentzian wormholes
  in Galileon theories with an extra scalar field},''
  \href{http://dx.doi.org/10.1103/PhysRevD.94.123516}{{\em Phys. Rev.}
  {\bfseries D94} no.~12, (2016) 123516},
\href{http://arxiv.org/abs/1607.04099}{{\ttfamily arXiv:1607.04099 [hep-th]}}.

\bibitem{Creminelli:2016zwa}
P.~Creminelli, D.~Pirtskhalava, L.~Santoni, and E.~Trincherini, ``{Stability of
  Geodesically Complete Cosmologies},''
  \href{http://dx.doi.org/10.1088/1475-7516/2016/11/047}{{\em JCAP} {\bfseries
  1611} no.~11, (2016) 047},
\href{http://arxiv.org/abs/1610.04207}{{\ttfamily arXiv:1610.04207 [hep-th]}}.

\bibitem{Cai:2016thi}
Y.~Cai, Y.~Wan, H.-G. Li, T.~Qiu, and Y.-S. Piao, ``{The Effective Field Theory
  of nonsingular cosmology},''
  \href{http://dx.doi.org/10.1007/JHEP01(2017)090}{{\em JHEP} {\bfseries 01}
  (2017) 090},
\href{http://arxiv.org/abs/1610.03400}{{\ttfamily arXiv:1610.03400 [gr-qc]}}.

\bibitem{Hawking:1969sw}
S.~W. Hawking and R.~Penrose, ``{The Singularities of gravitational collapse
  and cosmology},''
\href{http://dx.doi.org/10.1098/rspa.1970.0021}{{\em Proc. Roy. Soc. Lond.}
  {\bfseries A314} (1970) 529--548}.

\bibitem{Creminelli:2012my}
P.~Creminelli, K.~Hinterbichler, J.~Khoury, A.~Nicolis, and E.~Trincherini,
  ``{Subluminal Galilean Genesis},''
  \href{http://dx.doi.org/10.1007/JHEP02(2013)006}{{\em JHEP} {\bfseries 1302}
  (2013) 006},
\href{http://arxiv.org/abs/1209.3768}{{\ttfamily arXiv:1209.3768 [hep-th]}}.

\bibitem{Hinterbichler:2012fr}
K.~Hinterbichler, A.~Joyce, J.~Khoury, and G.~E.~J. Miller, ``{DBI Realizations
  of the Pseudo-Conformal Universe and Galilean Genesis Scenarios},''
  \href{http://dx.doi.org/10.1088/1475-7516/2012/12/030}{{\em JCAP} {\bfseries
  1212} (2012) 030},
\href{http://arxiv.org/abs/1209.5742}{{\ttfamily arXiv:1209.5742 [hep-th]}}.

\bibitem{Hinterbichler:2012yn}
K.~Hinterbichler, A.~Joyce, J.~Khoury, and G.~E.~J. Miller,
  ``{Dirac-Born-Infeld Genesis: An Improved Violation of the Null Energy
  Condition},'' \href{http://dx.doi.org/10.1103/PhysRevLett.110.241303}{{\em
  Phys. Rev. Lett.} {\bfseries 110} no.~24, (2013) 241303},
\href{http://arxiv.org/abs/1212.3607}{{\ttfamily arXiv:1212.3607 [hep-th]}}.

\bibitem{Pirtskhalava:2014esa}
D.~Pirtskhalava, L.~Santoni, E.~Trincherini, and P.~Uttayarat, ``{Inflation
  from Minkowski Space},''
  \href{http://dx.doi.org/10.1007/JHEP12(2014)151}{{\em JHEP} {\bfseries 12}
  (2014) 151},
\href{http://arxiv.org/abs/1410.0882}{{\ttfamily arXiv:1410.0882 [hep-th]}}.

\bibitem{Kobayashi:2015gga}
T.~Kobayashi, M.~Yamaguchi, and J.~Yokoyama, ``{Galilean Creation of the
  Inflationary Universe},''
  \href{http://dx.doi.org/10.1088/1475-7516/2015/07/017}{{\em JCAP} {\bfseries
  1507} no.~07, (2015) 017},
\href{http://arxiv.org/abs/1504.05710}{{\ttfamily arXiv:1504.05710 [hep-th]}}.

\bibitem{Nishi:2015pta}
S.~Nishi and T.~Kobayashi, ``{Generalized Galilean Genesis},''
  \href{http://dx.doi.org/10.1088/1475-7516/2015/03/057}{{\em JCAP} {\bfseries
  1503} no.~03, (2015) 057},
\href{http://arxiv.org/abs/1501.02553}{{\ttfamily arXiv:1501.02553 [hep-th]}}.

\bibitem{Brandenberger:2016vhg}
R.~Brandenberger and P.~Peter, ``{Bouncing Cosmologies: Progress and
  Problems},'' \href{http://dx.doi.org/10.1007/s10701-016-0057-0}{{\em Found.
  Phys.} {\bfseries 47} no.~6, (2017) 797--850},
\href{http://arxiv.org/abs/1603.05834}{{\ttfamily arXiv:1603.05834 [hep-th]}}.

\bibitem{Battefeld:2014uga}
D.~Battefeld and P.~Peter, ``{A Critical Review of Classical Bouncing
  Cosmologies},'' \href{http://dx.doi.org/10.1016/j.physrep.2014.12.004}{{\em
  Phys. Rept.} {\bfseries 571} (2015) 1--66},
\href{http://arxiv.org/abs/1406.2790}{{\ttfamily arXiv:1406.2790
  [astro-ph.CO]}}.

\bibitem{Rubakov:2013kaa}
V.~A. Rubakov, ``{Consistent NEC-violation: towards creating a universe in the
  laboratory},'' \href{http://dx.doi.org/10.1103/PhysRevD.88.044015}{{\em Phys.
  Rev.} {\bfseries D88} (2013) 044015},
\href{http://arxiv.org/abs/1305.2614}{{\ttfamily arXiv:1305.2614 [hep-th]}}.

\bibitem{Adams:2006sv}
A.~Adams, N.~Arkani-Hamed, S.~Dubovsky, A.~Nicolis, and R.~Rattazzi,
  ``{Causality, analyticity and an IR obstruction to UV completion},''
  \href{http://dx.doi.org/10.1088/1126-6708/2006/10/014}{{\em JHEP} {\bfseries
  0610} (2006) 014},
\href{http://arxiv.org/abs/hep-th/0602178}{{\ttfamily arXiv:hep-th/0602178
  [hep-th]}}.

\bibitem{Evslin:2011rj}
J.~Evslin, ``{Stability of Closed Timelike Curves in a Galileon Model},''
  \href{http://dx.doi.org/10.1007/JHEP03(2012)009}{{\em JHEP} {\bfseries 03}
  (2012) 009},
\href{http://arxiv.org/abs/1112.1349}{{\ttfamily arXiv:1112.1349 [hep-th]}}.

\bibitem{Easson:2013bda}
D.~A. Easson, I.~Sawicki, and A.~Vikman, ``{When Matter Matters},''
  \href{http://dx.doi.org/10.1088/1475-7516/2013/07/014}{{\em JCAP} {\bfseries
  1307} (2013) 014},
\href{http://arxiv.org/abs/1304.3903}{{\ttfamily arXiv:1304.3903 [hep-th]}}.

\bibitem{Babichev:2007dw}
E.~Babichev, V.~Mukhanov, and A.~Vikman, ``{k-Essence, superluminal
  propagation, causality and emergent geometry},''
  \href{http://dx.doi.org/10.1088/1126-6708/2008/02/101}{{\em JHEP} {\bfseries
  02} (2008) 101},
\href{http://arxiv.org/abs/0708.0561}{{\ttfamily arXiv:0708.0561 [hep-th]}}.

\bibitem{Geroch:2010da}
R.~Geroch, ``{Faster Than Light?},'' {\em AMS/IP Stud. Adv. Math.} {\bfseries
  49} (2011) 59--70,
\href{http://arxiv.org/abs/1005.1614}{{\ttfamily arXiv:1005.1614 [gr-qc]}}.

\bibitem{Bruneton:2006gf}
J.-P. Bruneton, ``{On causality and superluminal behavior in classical field
  theories. Applications to k-essence theories and MOND-like theories of
  gravity},'' \href{http://dx.doi.org/10.1103/PhysRevD.75.085013}{{\em Phys.
  Rev.} {\bfseries D75} (2007) 085013},
\href{http://arxiv.org/abs/gr-qc/0607055}{{\ttfamily arXiv:gr-qc/0607055}}.

\bibitem{ArmendarizPicon:2005nz}
C.~Armendariz-Picon and E.~A. Lim, ``{Haloes of k-essence},''
  \href{http://dx.doi.org/10.1088/1475-7516/2005/08/007}{{\em JCAP} {\bfseries
  0508} (2005) 007}, \href{http://arxiv.org/abs/astro-ph/0505207}{{\ttfamily
  arXiv:astro-ph/0505207 [astro-ph]}}.

\bibitem{Kang:2007vs}
J.~U. Kang, V.~Vanchurin, and S.~Winitzki, ``{Attractor scenarios and
  superluminal signals in k-essence cosmology},''
  \href{http://dx.doi.org/10.1103/PhysRevD.76.083511}{{\em Phys. Rev.}
  {\bfseries D76} (2007) 083511},
\href{http://arxiv.org/abs/0706.3994}{{\ttfamily arXiv:0706.3994 [gr-qc]}}.

\bibitem{Bruneton:2007si}
J.-P. Bruneton and G.~Esposito-Farese, ``{Field-theoretical formulations of
  MOND-like gravity},''
  \href{http://dx.doi.org/10.1103/PhysRevD.76.124012}{{\em Phys. Rev.}
  {\bfseries D76} (2007) 124012},
\href{http://arxiv.org/abs/0705.4043}{{\ttfamily arXiv:0705.4043 [gr-qc]}}.

\bibitem{Evslin:2011vh}
J.~Evslin and T.~Qiu, ``{Closed Timelike Curves in the Galileon Model},''
  \href{http://dx.doi.org/10.1007/JHEP11(2011)032}{{\em JHEP} {\bfseries 11}
  (2011) 032},
\href{http://arxiv.org/abs/1106.0570}{{\ttfamily arXiv:1106.0570 [hep-th]}}.

\bibitem{Hawking:1991nk}
S.~W. Hawking, ``{The Chronology protection conjecture},''
\href{http://dx.doi.org/10.1103/PhysRevD.46.603}{{\em Phys. Rev.} {\bfseries
  D46} (1992) 603--611}.

\bibitem{Burrage:2011cr}
C.~Burrage, C.~de~Rham, L.~Heisenberg, and A.~J. Tolley, ``{Chronology
  Protection in Galileon Models and Massive Gravity},''
  \href{http://dx.doi.org/10.1088/1475-7516/2012/07/004}{{\em JCAP} {\bfseries
  1207} (2012) 004},
\href{http://arxiv.org/abs/1111.5549}{{\ttfamily arXiv:1111.5549 [hep-th]}}.

\bibitem{Dvali:2010jz}
G.~Dvali, G.~F. Giudice, C.~Gomez, and A.~Kehagias, ``{UV-Completion by
  Classicalization},'' \href{http://dx.doi.org/10.1007/JHEP08(2011)108}{{\em
  JHEP} {\bfseries 1108} (2011) 108},
\href{http://arxiv.org/abs/1010.1415}{{\ttfamily arXiv:1010.1415 [hep-ph]}}.

\bibitem{Dvali:2010ns}
G.~Dvali and D.~Pirtskhalava, ``{Dynamics of Unitarization by
  Classicalization},''
  \href{http://dx.doi.org/10.1016/j.physletb.2011.03.054}{{\em Phys.Lett.}
  {\bfseries B699} (2011) 78--86},
\href{http://arxiv.org/abs/1011.0114}{{\ttfamily arXiv:1011.0114 [hep-ph]}}.

\bibitem{Dvali:2016ovn}
G.~Dvali, ``{Strong Coupling and Classicalization},''
  \href{http://dx.doi.org/10.1142/9789813208292_0005}{{\em Subnucl. Ser.}
  {\bfseries 53} (2017) 189--200},
\href{http://arxiv.org/abs/1607.07422}{{\ttfamily arXiv:1607.07422 [hep-th]}}.

\bibitem{Vikman:2012bx}
A.~Vikman, ``{Suppressing Quantum Fluctuations in Classicalization},''
  \href{http://dx.doi.org/10.1209/0295-5075/101/34001}{{\em EPL} {\bfseries
  101} no.~3, (2013) 34001},
\href{http://arxiv.org/abs/1208.3647}{{\ttfamily arXiv:1208.3647 [hep-th]}}.

\bibitem{Dvali:2011nj}
G.~Dvali, ``{Classicalize or not to Classicalize?},''
\href{http://arxiv.org/abs/1101.2661}{{\ttfamily arXiv:1101.2661 [hep-th]}}.

\bibitem{Dvali:2012zc}
G.~Dvali, A.~Franca, and C.~Gomez, ``{Road Signs for UV-Completion},''
\href{http://arxiv.org/abs/1204.6388}{{\ttfamily arXiv:1204.6388 [hep-th]}}.

\bibitem{Keltner:2015xda}
L.~Keltner and A.~J. Tolley, ``{UV properties of Galileons: Spectral
  Densities},''
\href{http://arxiv.org/abs/1502.05706}{{\ttfamily arXiv:1502.05706 [hep-th]}}.

\bibitem{Dvali:2007hz}
G.~Dvali, ``{Black Holes and Large N Species Solution to the Hierarchy
  Problem},'' \href{http://dx.doi.org/10.1002/prop.201000009}{{\em Fortsch.
  Phys.} {\bfseries 58} (2010) 528--536},
\href{http://arxiv.org/abs/0706.2050}{{\ttfamily arXiv:0706.2050 [hep-th]}}.

\bibitem{Dubovsky:2005xd}
S.~Dubovsky, T.~Gregoire, A.~Nicolis, and R.~Rattazzi, ``{Null energy condition
  and superluminal propagation},''
  \href{http://dx.doi.org/10.1088/1126-6708/2006/03/025}{{\em JHEP} {\bfseries
  0603} (2006) 025}, \href{http://arxiv.org/abs/hep-th/0512260}{{\ttfamily
  arXiv:hep-th/0512260 [hep-th]}}.

\bibitem{Quintin:2015rta}
J.~Quintin, Z.~Sherkatghanad, Y.-F. Cai, and R.~H. Brandenberger, ``{Evolution
  of cosmological perturbations and the production of non-Gaussianities through
  a nonsingular bounce: Indications for a no-go theorem in single field matter
  bounce cosmologies},''
  \href{http://dx.doi.org/10.1103/PhysRevD.92.063532}{{\em Phys. Rev.}
  {\bfseries D92} no.~6, (2015) 063532},
\href{http://arxiv.org/abs/1508.04141}{{\ttfamily arXiv:1508.04141 [hep-th]}}.

\bibitem{Battarra:2014tga}
L.~Battarra, M.~Koehn, J.-L. Lehners, and B.~A. Ovrut, ``{Cosmological
  Perturbations Through a Non-Singular Ghost-Condensate/Galileon Bounce},''
  \href{http://dx.doi.org/10.1088/1475-7516/2014/07/007}{{\em JCAP} {\bfseries
  1407} (2014) 007},
\href{http://arxiv.org/abs/1404.5067}{{\ttfamily arXiv:1404.5067 [hep-th]}}.

\bibitem{Ijjas:2017pei}
A.~Ijjas, ``{Space-time slicing in Horndeski theories and its implications for
  non-singular bouncing solutions},''
\href{http://arxiv.org/abs/1710.05990}{{\ttfamily arXiv:1710.05990 [gr-qc]}}.

\bibitem{Felder:2002jk}
G.~N. Felder, A.~V. Frolov, L.~Kofman, and A.~D. Linde, ``{Cosmology with
  negative potentials},''
  \href{http://dx.doi.org/10.1103/PhysRevD.66.023507}{{\em Phys. Rev.}
  {\bfseries D66} (2002) 023507},
\href{http://arxiv.org/abs/hep-th/0202017}{{\ttfamily arXiv:hep-th/0202017}}.

\bibitem{Mukhanov:2005sc}
V.~Mukhanov, {\em {Physical Foundations of Cosmology}}.
\newblock Cambridge University Press, 2005.
\newblock
\url{http://www-spires.fnal.gov/spires/find/books/www?cl=QB981.M89::2005}.
\newblock

\bibitem{Kolevatov:2017dze}
R.~Kolevatov, S.~Mironov, V.~Rubakov, N.~Sukhov, and V.~Volkova,
  ``{Perturbations in generalized Galileon theories},''
  \href{http://dx.doi.org/10.1103/PhysRevD.96.125012}{{\em Phys. Rev.}
  {\bfseries D96} no.~12, (2017) 125012},
\href{http://arxiv.org/abs/1708.04262}{{\ttfamily arXiv:1708.04262 [hep-th]}}.

\bibitem{Rubakov:2015gza}
V.~A. Rubakov, ``{Can Galileons support Lorentzian wormholes?},''
  \href{http://dx.doi.org/10.1134/S004057791605010X}{{\em Theor. Math. Phys.}
  {\bfseries 187} no.~2, (2016) 743--752},
\href{http://arxiv.org/abs/1509.08808}{{\ttfamily arXiv:1509.08808 [hep-th]}}.

\bibitem{Rubakov:2016zah}
V.~A. Rubakov, ``{More about wormholes in generalized Galileon theories},''
  \href{http://dx.doi.org/10.1134/S0040577916080080}{{\em Theor. Math. Phys.}
  {\bfseries 188} no.~2, (2016) 1253--1258},
  \href{http://arxiv.org/abs/1601.06566}{{\ttfamily arXiv:1601.06566
  [hep-th]}}.
[Teor. Mat. Fiz.188,no.2,337(2016)].

\bibitem{Evseev:2016ppw}
O.~A. Evseev and O.~I. Melichev, ``{Instability of static semiclosed worlds in
  generalized Galileon theories},''
  \href{http://dx.doi.org/10.1103/PhysRevD.96.024030}{{\em Phys. Rev.}
  {\bfseries D96} no.~2, (2017) 024030},
\href{http://arxiv.org/abs/1607.01721}{{\ttfamily arXiv:1607.01721 [hep-th]}}.

\end{thebibliography}\endgroup

\end{document}